\documentclass[prd,onecolumn,nopacs,nofootinbib]{revtex4}

\usepackage{amsmath}
\usepackage{graphicx}
\usepackage{amsfonts}
\usepackage{latexsym}
\usepackage{bbold}
\usepackage{wasysym}
\usepackage{calligra}
\usepackage{float}
\usepackage{ulem}
\usepackage{inputenc}
\usepackage{xspace}
\usepackage{url}
\usepackage{epstopdf}
\usepackage{tikz}

\usepackage{framed} 
\usepackage[framed]{ntheorem}
\newframedtheorem{frm-thm}{Theorem}

\newcommand{\be}{\begin{equation}}
\newcommand{\ee}{\end{equation}}
\newcommand{\beq} {\begin{equation}}
\newcommand{\eeq} {\end{equation}}
\newcommand{\ba}{\begin{eqnarray}}
\newcommand{\ea}{\end{eqnarray}}

\begin{document}

\title{Cosmology of Quadratic Metric-Affine Gravity}
	
\author{Damianos Iosifidis$^\ast$, Lucrezia Ravera$^{\dag,\star}$}

\affiliation{$^\ast$Institute of Theoretical Physics, Department of Physics, Aristotle University of Thessaloniki, 54124 Thessaloniki, Greece. \\
			$^\dag$DISAT, Politecnico di Torino, Corso Duca degli Abruzzi 24, 10129 Torino, Italy. \\
			$^\star$INFN, Sezione di Torino, Via P. Giuria 1, 10125 Torino, Italy.}

\email{diosifid@auth.gr, lucrezia.ravera@polito.it}
	
\date{\today}

\begin{abstract}
		
We investigate the cosmological aspects of the most general parity preserving Metric-Affine Gravity theory quadratic in torsion and non-metricity in the presence of a cosmological hyperfluid. The equations of motion are obtained by varying the action with respect to the metric and the independent affine connection. Subsequently, considering a Friedmann-Lemaître-Robertson-Walker background, we derive the most general form of the modified Friedmann equations for the full quadratic theory. We then focus on a characteristic sub-case involving only two quadratic contributions given in terms of torsion and non-metricity vectors. In this setup, studying the modified Friedmann equations along with the conservation laws of the perfect cosmological hyperfluid, we provide exact solutions both for purely dilation and for purely spin hypermomentum sources. We then discuss the physical consequences of our model and the prominent role of torsion and non-metricity in this cosmological setup.

\end{abstract}
	
\maketitle
	
\allowdisplaybreaks
	
	
\tableofcontents

\section{Introduction}\label{intro}

Modified gravity models \cite{CANTATA:2021ktz} beyond General Relativity (GR) have received considerable attention in recent years, as they might provide solutions to both small and large scales open issues that still afflicts Einstein's well-celebrated theory of gravity.
GR is based on Riemannian geometry and, on the other hand, proposed generalizations and extensions of GR have they roots in non-Riemannian geometry \cite{Eisenhart:1927}, where the assumptions of metric compatibility and torsionlessness of the affine connection are released. In other words, in this framework non-vanishing torsion and non-metricity are allowed both having a clear geometric interpretation. The former ensures that no infinitesimal parallelogram can be formed (i.e., they crack into pentagons), while the latter causes dot products and lengths of vectors to change under parallel transport in spacetime.

A prominent class of theories developed in the non-Riemannian setup goes under the name of Metric-Affine Gravity (MAG) \cite{Hehl:1994ue,Hehl:1999sb,Iosifidis:2019jgi}. The latter can be formulated as a ``gauge'' theory of gravity \cite{Hehl:1994ue,Hehl:1999sb}, but this is not mandatory, as the development of gravitational models out of the gauge realm already turned out to be relevant by itself (cf, e.g., \cite{Iosifidis:2019jgi,Klemm:2020mfp,Klemm:2020gfm,Iosifidis:2020dck}). In this context, the metric and the connection can be treated as independent objects without assuming any form or symmetry or compatibility condition for the general affine connection from the very beginning. The latter is eventually obtained from the study of the field equations derived in the first order (i.e., Palatini) formalism.\footnote{The literature on the topic is huge. For a recent review of Einstein manifolds with torsion and non-metricity see, e.g., \cite{Klemm:2018bil}. Besides, let us also mention that Einstein-Cartan-Weyl spaces in three dimensions, involving torsion and non-metricity, have recently proved to describe the near-horizion geometry of some supergravity BPS black holes solutions \cite{Klemm:2019izb}. Furthermore, the  ($3+1$)-formulation for gravity with torsion and non-metricity has been recently developed in \cite{Ariwahjoedi:2020wmo,Ariwahjoedi:2021yth}.} Another characteristic feature of MAG is the fact that the matter Lagrangian is in principle allowed to depend on the general affine connection as well. The coupling is encoded into the so-called hypermomentum tensor, which encompasses the microstructure of matter \cite{Hehl:1994ue,Hehl:1976kt,Hehl:1976kt2}. In particular, it describes dilation, spin, and shear sources. This last intriguing characteristic of MAG, namely the microstructure-extended geometry interrelation, makes it especially interesting.

Under the perspective described above, MAG theories result to be rather general and powerful \cite{Iosifidis:2018diy,Iosifidis:2018jwu,Capozziello:2009mq,Percacci:2020ddy,BeltranJimenez:2020sqf,Aoki:2019rvi,Cabral:2020fax,Ariwahjoedi:2020wmo,Ariwahjoedi:2021yth,Helpin:2019vrv,Bahamonde:2020fnq}, and can possibly lead to diverse scenarios beyond GR, which may have relevant consequences, for instance, in the cosmological context. For some relevant results already obtained in MAG cosmology we refer the reader to, e.g., \cite{BeltranJimenez:2015pnp,BeltranJimenez:2017vop,Kranas:2018jdc,Barragan:2009sq,Shimada:2018lnm,Kubota:2020ehu,Mikura:2020qhc,Mikura:2021ldx,Bombacigno:2018tyw,Bombacigno:2021bpk,Iosifidis:2021crj}. In particular, in this paper we are interested in the cosmological aspects, in a Friedmann-Lema\^{i}tre-Robertson-Walker (FLRW) background, of the most general MAG theory in $n$ spacetime dimensions involving all possible parity preserving quadratic terms in torsion and non-metricity, in the presence of a cosmological hyperfluid. The latter is a classical continuous medium carrying hypermomentum \cite{Obukhov:1993pt,Obukhov:1996mg,Babourova:1998mgh,Ray:1982qr,Smalley}. A novel, general formulation of perfect hyperfluid generalizing the classical perfect fluid notion of GR has been recently presented in \cite{Iosifidis:2020gth,Iosifidis:2021nra,Iosifidis:2020upr} by first giving its physical definition and later using the appropriate mathematical formulation in order to extract its energy tensors by demanding spatial isotropy. We will refer to the aforementioned perfect cosmological hyperfluid in this work.

Quadratic MAG theories have been previously considered in \cite{Tucker:1995fw,Obukhov:1996pf,Vlachynsky:1996zh,Dereli:1996ex,Obukhov:1997zd,Vitagliano:2010sr,Baekler:2011jt,Vitagliano:2013rna,Benisty:2018ufz,Benisty:2021sul}. More precisely, in \cite{Obukhov:1997zd} a quadratic theory involving also a piece quadratic in the so-called homothetic curvature tensor was studied, while in \cite{Vitagliano:2010sr} only contributions quadratic in the torsion tensor were taken into account. In the cosmological context, in \cite{Iosifidis:2021iuw} the cosmology of a quadratic metric-compatible torsionful gravity theory in the presence of a perfect hyperfluid was presented and analyzed for the first time. The current work also provides a generalization of the results presented in \cite{Iosifidis:2021iuw} to the case in which both torsion and non-metricity are non-vanishing.

The remaining of this paper is structured as follows: In Section \ref{tb} we review the theoretical background of MAG, cosmological aspects of torsion and non-metricity, and the perfect cosmological hyperfluid. In Section \ref{thethy} we study the most general parity preserving MAG theory quadratic in torsion and non-metricity in the presence of a perfect hyperfluid. The gravitational part of the action is given by the (non-Riemannian) Einstein-Hilbert contribution plus all possible parity preserving torsion and non-metricity scalars. We derive the field equations in the first order formalism, treating the metric and the general affine connection as independent variables. Section \ref{cosmquadrMAG} is devoted to the study of the cosmology of the quadratic theory. First of all, we consider a FLRW background and obtain the modified Friedmann equations of the general quadratic theory. Then, we focus on a characteristic sub-case involving only two quadratic contributions given in terms of torsion and non-metricity vectors and, in this setup, from the study of the modified Friedmann equations and the conservation laws of the perfect cosmological hyperfluid, we are able to provide exact solutions in the cases of purely dilation and purely spin hypermomentum types. We conclude our work with some remarks and possible future developments. Useful formulas are collected in Appendix \ref{appa}.

\section{Theoretical background}\label{tb}

In this section we briefly give all the theoretical background necessary for a clearer understanding of the present work.

\subsection{Geometric setup}

Let us first review the geometric setup. We consider the framework of non-Riemannian geometry, and accordingly introduce on the manifold a metric tensor $g_{\mu \nu}$ and a general affine connection $\nabla$ with coefficients ${\Gamma^\lambda}_{\mu \nu}$. These geometric objects will be treated as independent, a priori. We consider $n$ spacetime dimensions and our convention for the metric signature is mostly plus. The generic decomposition of a general affine connection is
\begin{equation}
{\Gamma^\lambda}_{\mu \nu} = \tilde{\Gamma}^\lambda_{\phantom{\lambda} \mu \nu} + {N^\lambda}_{\mu \nu}\,,
\end{equation}
where
\beq\label{lcconn}
\tilde{\Gamma}^\lambda_{\phantom{\lambda}\mu\nu} = \frac12 g^{\rho\lambda}\left(\partial_\mu 
g_{\nu\rho} + \partial_\nu g_{\rho\mu} - \partial_\rho g_{\mu\nu}\right)\,.
\eeq
is the Levi-Civita connection and 
\beq\label{distortion}
{N^\lambda}_{\mu\nu} = \underbrace{\frac12 g^{\rho\lambda}\left(Q_{\mu\nu\rho} + Q_{\nu\rho\mu}
- Q_{\rho\mu\nu}\right)}_{\text{deflection {(or disformation)}}} - \underbrace{g^{\rho\lambda}\left(S_{\rho\mu\nu} +
S_{\rho\nu\mu} - S_{\mu\nu\rho}\right)}_{\text{contorsion} \, := \, {K^\lambda}_{\mu \nu}}\,,
\eeq
is the distortion tensor, given in terms of torsion,
\beq
{S_{\mu\nu}}^\lambda := {\Gamma^\lambda}_{[\mu\nu]}\,, \quad S_{\mu \nu \alpha} = N_{\alpha[\mu \nu]} \,,\label{torsdef}
\eeq
and non-metricity
\beq
Q_{\lambda\mu\nu}:= -\nabla_\lambda g_{\mu\nu} = 
-\partial_\lambda g_{\mu\nu} + {\Gamma^\rho}_{\mu\lambda} g_{\rho\nu} +
{\Gamma^\rho}_{\nu\lambda}g_{\mu\rho} \,, \quad Q_{\nu \alpha \mu} = 2 N_{(\alpha \mu )\nu} \,. \label{nonmetdef}
\eeq
Both torsion and non-metricity have trace parts: $S_\lambda :={S_{\lambda \sigma}}^{\sigma}$ is the torsion vector, $Q_\lambda := {Q_{\lambda \mu}}^\mu$ and $q_\nu := {Q^\mu}_{\mu\nu}$ are the non-metricity vectors, wit the former oftentimes referred to as the Weyl vector. Furthermore, in the case $n=4$ one can also define the torsion pseudo-vector according to: $t^\rho := \epsilon^{\rho \lambda \mu \nu} S_{\lambda \mu \nu}$, where $\epsilon^{\rho \lambda \mu \nu}$ is the Levi-Civita tensor in four dimensions.
The covariant derivative $\nabla$ (associated with $\Gamma$) of, e.g., a vector $v^\lambda$ is given by
\begin{equation}
\nabla_\nu v^\lambda = \partial_\nu v^\lambda  + {\Gamma^\lambda}_{\mu \nu} v^\mu \,.
\end{equation}
We define the curvature (Riemann) tensor as
\beq\label{curvtensdef}
{R^\mu}_{\nu \alpha \beta} := 2 \partial_{[\alpha} {\Gamma^\mu}_{|\nu|\beta]} + 2 {\Gamma^\mu}_{\rho[\alpha} {\Gamma^\rho}_{|\nu|\beta]} = \tilde{R}^\mu_{\phantom{\mu} \nu \alpha \beta} + 2 \tilde{\nabla}_{[\alpha} {N^\mu}_{|\nu|\beta]} + 2 {N^\mu}_{\lambda|\alpha} {N^\lambda}_{|\nu|\beta]} \,,
\eeq
where $\tilde{\nabla}$ denotes the Levi-Civita covariant derivative.
In particular, one can form the following contractions of the curvature tensor ${R^\mu}_{\nu \alpha \beta}$:
\begin{align}
& R_{\nu \beta} := {R^\mu}_{\nu \mu \beta} \,, \label{Riccitens} \\
& \hat{R}_{\alpha \beta} := {R^\mu}_{\mu \alpha \beta} = \partial_{[\alpha} Q_{\beta]} \,, \label{homot} \\
& \check{R}^\lambda_{\phantom{\lambda} \alpha} := {R^\lambda}_{\mu \nu \alpha} g^{\mu \nu} \,, \label{Rcheck}
\end{align}
where \eqref{Riccitens} defines the Ricci tensor of $\Gamma$, \eqref{homot} the so-called homothetic curvature tensor, and \eqref{Rcheck} a third tensor referred to as the co-Ricci tensor in the literature. Finally, 
\begin{equation}
R:= R_{\mu \nu} g^{\mu \nu} = - \check{R}_{\mu \nu} g^{\mu \nu}
\end{equation}
is the curvature scalar associated with $\Gamma$ which is still uniquely defined. Note that the latter can be decomposed into its Riemannian part plus non-Riemannian contributions associated with torsion and non-metricity. Having introduced the minimum theoretical setup for the formalism we are going to be using we can now move on to the physics part and elaborate a little bit on the sources of MAG.

\subsection{Hypermomentum and energy-momentum tensors}

Let us now recall the concepts of energy-momentum and hypermomentum tensors, following \cite{Iosifidis:2020gth}. We will assume the full action to be a functional of the metric (and its derivatives), the general affine connection  and the matter fields (here collectively denoted by $\varphi$), that is
\beq \label{magact}
S[g,\Gamma,\varphi] = S_{\text{G}}[g,\Gamma] + S_{\text{M}}[g,\Gamma,\varphi] \,.
\eeq
In particular,
\beq
S_{\text{G}}[g,\Gamma] = \frac{1}{2\kappa} \int d^n x \sqrt{-g} \mathcal{L}_{\text{G}} (g,\Gamma) 
\eeq
is the gravitational part of the action ($\kappa=8\pi G$ is the gravitational constant) and
\beq
S_{\text{M}}[g,\Gamma,\varphi] = \int d^n x \sqrt{-g} \mathcal{L}_{\text{M}} (g,\Gamma,\varphi) 
\eeq
the matter part.
One can then define the metric energy-momentum tensor (MEMT, for short),
\beq
T_{\mu \nu} := - \frac{2}{\sqrt{-g}} \frac{\delta S_{\text{M}}}{\delta g^{\mu \nu}} = - \frac{2}{\sqrt{-g}} \frac{\delta (\sqrt{-g} \mathcal{L}_{\text{M}})}{\delta g^{\mu \nu}} \,,
\eeq
and the hypermomentum tensor \cite{Hehl:1994ue,Hehl:1976kt,Hehl:1976kt2},
\beq
{\Delta_\lambda}^{\mu \nu} := - \frac{2}{\sqrt{-g}} \frac{\delta S_{\text{M}}}{\delta {\Gamma^\lambda}_{\mu \nu}} = - \frac{2}{\sqrt{-g}} \frac{\delta (\sqrt{-g} \mathcal{L}_{\text{M}})}{\delta {\Gamma^\lambda}_{\mu \nu}} \,.
\eeq
On the other hand, working in the equivalent formalism based on the vielbeins ${e_{\mu}}^c$ and spin connection $\omega_{\mu | a b}$, where $a,b,\ldots$ are Lorentz (i.e., tangent) indices,\footnote{Here we have the usual relation $g_{\mu \nu}= {e_\mu}^a {e_\nu}^b \eta_{ab}$ connecting metric and vielbeins, being $\eta_{ab}$ the tangent space flat Minkowski metric.} one may also introduce the canonical energy-momentum tensor (CEMT),
\beq\label{cemt}
{t^\mu}_c := \frac{1}{\sqrt{-g}} \frac{\delta S_{\text{M}}}{\delta {e_\mu}^c} \,.
\eeq
Note that, while the MEMT is symmetric, the CEMT, in general, is not.
One can then prove that the CEMT is not independent of the metric energy-momentum and hypermomentum tensors \cite{Hehl:1994ue}, as the following relation holds  \cite{Iosifidis:2020gth}\footnote{In the explicit calculations one also exploits the identity $$\nabla_\nu {e_\mu}^a = 0 = \partial_\nu {e_\mu}^a - {\Gamma^\rho}_{\mu \nu} {e_\rho}^a + \omega^{\phantom{\nu} a}_{\nu \phantom{a} b} {e_\mu}^b \,,$$ which connects the two formalisms.}
\beq\label{cemt1}
{t^\mu}_\lambda := \frac{1}{\sqrt{-g}} \frac{\delta S_{\text{M}}}{\delta {e_\lambda}^c} {e_\lambda}^c = {T^\mu}_\lambda - \frac{1}{2 \sqrt{-g}} \hat{\nabla}_\nu \left( \sqrt{-g} {\Delta_\lambda}^{\mu \nu} \right) \,,
\eeq
where we have also defined a modified covariant derivative
\beq\label{hatnabla}
\hat{\nabla}_\nu := 2 S_\nu - \nabla_\nu \,.
\eeq
Observe that for matter such that $\Delta_{\alpha \mu \nu}\equiv 0$ (namely, for matter with no microstructure) the CEMT and MEMT coincide.
As we will also recall in the following, eq. \eqref{cemt1} turns out to be a conservation law of the cosmological hyperfluid \cite{Iosifidis:2020gth}.
Moreover, let us notice that from eq. \eqref{cemt1} one can derive the conservation law for spin \cite{Obukhov:2014nja},
\beq
2 t_{[\mu \nu]} = \frac{1}{\sqrt{-g}} \hat{\nabla}_\alpha \left( \sqrt{-g} {\tau_{\mu \nu}}^\alpha \right) - Q_{\alpha \beta [\mu} {\Delta_{\nu]}}^{\beta \alpha} \,,
\eeq
with
\beq
{\tau_{\mu \nu}}^\alpha := {\Delta_{[\mu \nu]}}^\alpha \,.
\eeq
As we can see, the conservation law for spin receives, in general, contributions from non-metricity. Let us also note here that in a homogeneous cosmological setup, any antisymmetric two-index object vanishes identically. This means that in a homogeneous cosmological setting $t_{[\mu \nu]}\equiv 0$ and, therefore, one is left only with the symmetric part of $t_{\mu \nu}$, that is $t_{\mu \nu}=t_{[\mu \nu]}$.
Finally, contracting eq. \eqref{cemt1} in $\mu,\lambda$, we find the following relation
\beq\label{tracerelation}
t = T + \frac{1}{2 \sqrt{-g}} \partial_\nu \left( \sqrt{-g} \Delta^\nu \right) \,,
\eeq
where
\beq
t := {t^\mu}_\mu \,, \quad T := {T^\mu}_\mu \,, \quad \Delta^\nu := {\Delta_\lambda}^{\lambda \nu} 
\eeq
are the CEMT and MEMT traces and a trace of the hypermomentum, respectively.
The above equation implies
\begin{align}
T = 0 \quad & \leftrightarrow \quad 2t = \frac{1}{\sqrt{-g}} \partial_\nu \left( \sqrt{-g} \Delta^\nu \right) \,, \\
t = 0 \quad & \leftrightarrow \quad 2T = - \frac{1}{\sqrt{-g}} \partial_\nu \left( \sqrt{-g} \Delta^\nu \right) \,, \\
t = T \quad & \leftrightarrow \quad \partial_\nu \left( \sqrt{-g} \Delta^\nu \right)  = 0 \,,
\end{align}
which can follow from considering specific matter and correspond, respectively, to conformally, frame rescalings, and special projective transformations invariant theories (cf. also \cite{Iosifidis:2018zwo}). From the latter equation it immediately follows that for Theories that respect the special projective symmetry\footnote{Also known as Einstein's $\lambda$ transformations.} the associated dilation current is always conserved. This is true for all classes of Theories whose gravitational part depends only on the Riemann tensor.

\subsection{Cosmological aspects of torsion and non-metricity}\label{cosmtornonmetsubsec}

In the following we recall key cosmological aspects of torsion and non-metricity, which will be useful in the reminder of this paper.
First of all, we will consider a homogeneous, flat (i.e., $K=0$, being $K$ a curvature parameter) FLRW spacetime with the usual Robertson-Walker line element
\beq\label{RWle}
ds^{2}=-dt^{2}+a^{2}\delta_{ij}dx^{i}dx^{j} \,,
\eeq
where $a(t)$ is the scale factor of the Universe and $i,j=1,2,\ldots,n-1$. In addition, we define the projector tensor
\beq\label{projop}
h_{\mu \nu}:= g_{\mu \nu} + u_\mu u_\nu \,,
\eeq
which project objects on the space orthogonal to $u^\mu$, the latter being the normalized $n$-velocity field of a given fluid which in co-moving coordinates is expressed as $u^\mu= \delta^\mu_0=(1,0,0,\ldots,0)$, $u_\mu u^\mu=-1$. Accordingly, let us also introduce the temporal derivative
\beq\label{tempdev}
\dot{}=u^\alpha \nabla_\alpha \,.
\eeq
The above constitutes a $1+(n-1)=1+m$ spacetime split.

\subsubsection{Torsion and non-metricity in FLRW spacetime}

In a non-Riemannian FLRW spacetime in $1+3$ dimensions the general affine connection can be written as \cite{Iosifidis:2020gth}
\begin{equation}\label{connFLRW}
{\Gamma^\lambda}_{\mu \nu} = \tilde{\Gamma}^\lambda_{\phantom{\lambda}\mu \nu} + X(t) u^\lambda h_{\mu \nu} + Y(t) u_\mu {h^\lambda}_\nu + Z(t) u_\nu {h^\lambda}_\mu + V(t) u^\lambda u_\mu u_\nu + {\epsilon^\lambda}_{\mu \nu \rho} u^\rho W(t) \,.
\end{equation}
In particular, the non-vanishing components of the Levi-Civita connection are
\begin{equation}
\tilde{\Gamma}^0_{\phantom{0}ij} = \tilde{\Gamma}^0_{\phantom{0}ji} = \dot{a} a \delta_{ij} = H g_{ij} \,, \quad \tilde{\Gamma}^i_{\phantom{i}j0} = \tilde{\Gamma}^i_{\phantom{i}0j} = \frac{\dot{a}}{a} \delta^i_j = H \delta^i_j \,,
\end{equation}
where $H:=\frac{\dot{a}}{a}$ is the Hubble parameter.
On the other hand, the torsion and non-metricity tensors can be written, respectively, in the following way \cite{Iosifidis:2020gth,Iosifidis:2020zzp}:
\begin{equation}\label{tornonmetFLRW}
\begin{split}
S^{(n)}_{\mu \nu \alpha} & = 2 u_{[\mu} h_{\nu]\alpha} \Phi(t) + \epsilon_{\mu \nu \alpha \rho} u^\rho P(t) \delta^n_4 \,, \\
Q_{\alpha \mu \nu} & = A(t) u_\alpha h_{\mu \nu} + B(t) h_{\alpha(\mu} u_{\nu)} + C(t) u_\alpha u_\mu u_\nu \,,
\end{split}
\end{equation}
where $\delta^n_4$ is the Kronecker's delta ($\delta^{n=4}_4=1$, otherwise it gives zero). Given \eqref{tornonmetFLRW}, we may also compute
\begin{equation}\label{tornmtracesFLRW}
\begin{split}
& S_\alpha = (n-1) \Phi u_\alpha \,, \\
& Q_\mu = (n-1) A u_\mu - C u_\mu \,, \\
& q_\mu = \frac{(n-1)}{2} B u_\mu - C u_\mu \,.
\end{split}
\end{equation}
The functions $X(t)$, $Y(t)$, $Z(t)$, $V(t)$, $W(t)$ in \eqref{connFLRW} and $\Phi(t)$, $P(t)$, $A(t)$, $B(t)$, $C(t)$ in \eqref{tornonmetFLRW} describe non-Riemannian cosmological effects and give, together with the scale factor, the cosmic evolution of non-Riemannian geometries.\footnote{Notice that $P(t)$ is a pseudo-scalar and $\epsilon_{\mu \nu \alpha \beta}u^\beta$ is, by definition, purely spatial.} 
Moreover, using the definition of a general affine connection we have previously introduced, one can prove that
\begin{equation}
2(X+Y) = B \,, \quad 2 Z = A \,, \quad 2 V = C \,, \quad 2 \Phi = Y - Z \,, \quad P=W \,.
\end{equation}
The latter can also be inverted to get
\begin{equation}
W = P \,, \quad V= \frac{C}{2} \,, \quad Z = \frac{A}{2} \,, \quad Y = 2 \Phi + \frac{A}{2} \,, \quad X = \frac{B}{2} - 2 \Phi - \frac{A}{2} \,.
\end{equation}
Here, let us also collect some useful formulas we have derived in the current cosmological context.

\vspace{0.2cm}

\paragraph{Torsion scalars.}

Using the first of \eqref{tornonmetFLRW}, we find the following relations:\footnote{ $P=P(t)$ is there only in $n=4$ ($n=1+3$, $m=3$) spacetime dimensions. For  $m\neq 3$ it identically vanishes, i.e. $P\equiv 0$.}
\begin{equation}\label{puretor}
\begin{split}
& S_{\mu\nu\alpha}S^{\mu\nu\alpha}=-2(n-1)\Phi^{2}+6 P^{2} \delta^n_4 \,, \\
& S_{\mu\nu\alpha}S^{\alpha\mu\nu}=(n-1)\Phi^{2}+6 P^{2} \delta^n_4 \,, \\
& S_{\mu}S^{\mu}=-(n-1)^{2}\Phi^{2} \,.
\end{split}
\end{equation}
Besides, one might introduce the torsion scalar $\mathcal{T}$, given by
\begin{equation}
\mathcal{T} := S_{\mu\nu\alpha}S^{\mu\nu\alpha}-2S_{\mu\nu\alpha}S^{\alpha\mu\nu}-4S_{\mu}S^{\mu} \,,
\end{equation}
and prove that 
\begin{equation}
\mathcal{T} =4(n-1)(n-2)\Phi^{2}-6P^{2} \delta^n_4
\end{equation}
in our cosmological setup.

\vspace{0.2cm}

\paragraph{Mixed torsion--non-metricity scalars.}

Regarding mixed terms, we obtain
\begin{equation}\label{mixed}
\begin{split}
& Q_{\mu\nu\alpha}S^{\mu\nu\alpha}=-(n-1) A\Phi +\frac{(n-1)}{2} B\Phi \,, \\
& Q_{\mu}S^{\mu}=(n-1)\Phi \Big[ C-(n-1)A \Big] \,, \\
& q_\mu S^{\mu}=(n-1) \Phi \Big[ C-\frac{(n-1)}{2}B \Big] \,.
\end{split}
\end{equation}
One could then define a $\mathcal{Q\ast T}$ mixed torsion--non-metricity scalar
\beq
\mathcal{Q\ast T} := 2 Q_{\alpha\mu\nu}S^{\alpha\mu\nu}+2 S_{\mu}(q^{\mu}-Q^{\mu}) 
\eeq
and prove that we have
\beq
\mathcal{Q\ast T} =(n-1)(n-2)\Phi (2 A-B) 
\eeq
in the current cosmological setting.

\vspace{0.2cm}

\paragraph{Non-metricity scalars.}

Concerning pure non-metricity scalars, we compute
\begin{equation}\label{purenonmet}
\begin{split}
& X_{1}:=Q_{\alpha\mu\nu}Q^{\alpha\mu\nu}=-\left[ (n-1)A^{2}+\frac{(n-1)}{2}B^{2}+C^{2} \right]  \,, \\
& X_{2}:=Q_{\alpha\mu\nu}Q^{\mu\nu\alpha}=-\Big[ (n-1)AB+\frac{(n-1)}{4} B^{2}+C^{2} \Big] \,, \\
& X_{3}:=Q_{\alpha}Q^{\alpha}=-\Big[ (n-1)A-C \Big]^{2} \,, \\
& X_{4}:=Q_{\alpha}q^{\alpha}=-\Big[ (n-1)A-C \Big]\Big[ \frac{(n-1)}{2}B -C \Big] \,, \\
& X_{5}:=q_{\alpha}q^{\alpha}=-\left[ \frac{(n-1)}{2}B-C \right]^{2} \,.
\end{split}
\end{equation}
Finally, defining 
\begin{equation}
\mathcal{Q} := \frac{1}{4} Q_{\alpha \mu \nu} Q^{\alpha \mu \nu} - \frac{1}{2} Q_{\alpha \mu \nu} Q^{\mu \nu \alpha} - \frac{1}{4} Q_\mu Q^\mu + \frac{1}{2} Q_\mu q^\mu \,,
\end{equation}
and, consequently, 
\begin{equation}
\mathcal{Z} := \mathcal{T} + \mathcal{Q} + \mathcal{Q\ast T} \,,
\end{equation}
we have that the Ricci scalar of $\Gamma$ can be written as
\beq
R=\tilde{R}+\mathcal{Z}+\tilde{\nabla}_{\mu}(q^{\mu}-Q^{\mu}-4S^{\mu}) \,.
\eeq
Then, as 
\beq
\tilde{\nabla}_{\mu}(q^{\mu}-Q^{\mu}-4S^{\mu})=(n-1)\frac{1}{\sqrt{-g}}\partial_{\mu} \left[ \sqrt{-g} u^{\mu}\left( \frac{B}{2}-A-4 \Phi \right) \right] \,,
\eeq
where we have used
\beq
\tilde{\nabla}_{\mu}(\sqrt{-g}\xi^{\mu})= \partial_{\mu}(\sqrt{-g}\xi^{\mu}) \,, \quad \forall \text{ vector } \xi^{\mu} \,,
\eeq
we find 
\begin{equation}
\begin{split}
R & = \tilde{R} + \frac{(n-1)}{4}\left[ (n-2)A^{2}- (n-3)AB+BC \right]+(n-1)(n-2)\Big( 4 \Phi^{2}+\Phi (2A-B)\Big) - 6 P^{2} \delta^n_4 \\
& +(n-1)\frac{1}{\sqrt{-g}}\partial_{\mu} \left[ \sqrt{-g} u^{\mu}\left( \frac{B}{2}-A-4 \Phi \right)    \right] \,,
\end{split}
\end{equation}
which is the final cosmological decomposition of the Ricci scalar of $\Gamma$ in terms of Riemannian (i.e., $\tilde{R}$) and non-Riemannian contributions. This decomposition is a key ingredient for the derivation of the modified Friedmann equations we are going to present in what follows.

\subsubsection{Relations among the invariants}\label{relinvsubs}

Note that in the highly symmetric FLRW background not all of the $11$ quadratic torsion and non-metricity invariants are independent.
Regarding the pure torsion invariants, since as we have seen above they involve only two combinations and we have three scalars (see \eqref{puretor}), the latter must be dependent. In fact, quite trivially we find
\beq
(n-1)S_{\mu\nu\alpha}S^{\mu\nu\alpha}-(n-1)S_{\mu\nu\alpha}S^{\alpha\mu\nu}-3 S_{\mu}S^{\mu}=0 \,.
\eeq
confirming that indeed only two out of the three pure torsion scalars are independent.
Analogously, concerning the mixed terms we observe that the difference of the last two in \eqref{mixed} is $(n-1)$ times the first one, namely
\beq
Q_{\mu}S^{\mu}-q_{\mu}S^{\mu}-(n-1)Q_{\mu\nu\alpha}S^{\mu\nu\alpha}=0 \,,
\eeq
indicating again their linear dependence. 
More tricky and complicated are the pure non-metricity scalars in \eqref{purenonmet}, as here we have $5$ expressions but $6$ combinations appearing. Then, considering their linear combinations, if the equation
\beq
\sum_{i=1}^{5}\lambda_{i}X_{i}=0
\eeq
has only the trivial solution $\lambda_{i}=0$ for all $i$, then the invariants are independent. If not all $\lambda$'s vanish, then the invariants will be dependent. In our case it happens that only $4$ equations are linearly independent, so we have a free parameter, which we take to be $\lambda_{3}$, and
\begin{equation}
\lambda_1 = - \lambda_3 (n-1) \,, \quad \lambda_2 = \lambda_3 (n-1) \,, \quad \lambda_4 = - 2 \lambda_3 \,, \quad \lambda_5 = \lambda_3 \,.
\end{equation}
Now, setting $\lambda_{3}=1$, the latter boil down to
\begin{equation}
\lambda_3 = 1 \,, \quad \lambda_{1}= 1-n \,, \quad \lambda_{2}=n-1 \,, \quad \lambda_{4}=-2 \,, \quad \lambda_{5}= 1 
\end{equation}
and we find that the non-metricity scalars obey the following constraint in $n$ spacetime dimensions:
\beq
(1-n) X_{1}+ (n-1)X_{2}+X_{3}-2X_{4}+X_{5}=0 \,,
\eeq
that is, only four out of the five non-metricity scalars are independent.
Note that in the case $n=4$ we get
\beq
\lambda_{1}=-3 \lambda_{3} \,, \quad \lambda_{2}=3 \lambda_{3} \,, \quad \lambda_{4}=-2 \lambda_{3} \,, \quad \lambda_{5}= \lambda_{3} \,,
\eeq
that is, setting $\lambda_3 = 1$,
\beq
\lambda_3 = 1 \,, \quad \lambda_{1}=-3 \,, \quad \lambda_{2}=3  \,, \quad \lambda_{4}=-2  \,, \quad \lambda_{5}=1 \,.
\eeq
Then, in $n=4$ one is left with
\beq
-3X_{1}+3X_{2}+X_{3}-2X_{4}+X_{5}=0 \,,
\eeq
 To recap, we have proved that in a cosmological background, out of the three torsion scalars only two are independent, out of the five non-metricity scalars four are independent and out of the mixed three torsion-non-metricity scalars two are independent. That is, in a homogeneous cosmological setting, only $8$ out of the $11$ parity even quadratic scalars in torsion and non-metricity are independent. This fact will be used later on in our analysis.

\subsection{Perfect cosmological hyperfluid}\label{perfhypsubs}

A hyperfluid is a classical continuous medium carrying hypermomentum. The general formulation of perfect cosmological hyperfluid generalizing the classical perfect fluid notion can be found in Ref. \cite{Iosifidis:2020gth,Iosifidis:2021nra}. In this paper we consider a perfect cosmological hyperfluid in a homogeneous cosmological setting, demanding also homogeneity.
The perfect hyperfluid is described in terms of the following MEMT and CEMT tensors \cite{Iosifidis:2020gth}
\beq\label{metrenmomform}
T_{\mu \nu} = \rho u_\mu u_\nu + p h_{\mu \nu} \,,
\eeq
and
\beq\label{canonenmomform}
t_{\mu \nu} = \rho_c u_\mu u_\nu + p_c h_{\mu \nu} \,,
\eeq
along with the hypermomentum tensor
\beq\label{hypermomform}
\Delta^{(n)}_{\alpha \mu \nu} = \phi(t) h_{\mu \alpha} u_\nu + \chi(t) h_{\nu \alpha} u_{\mu} + \psi(t) u_{\alpha} h_{\mu \nu} + \omega(t) u_\alpha u_\mu u_\nu + \delta^n_4 \epsilon_{\alpha \mu \nu \rho} u^\rho \zeta(t) \,,
\eeq
where $\rho$ and $p$ are the usual density and pressure of the perfect fluid component of the hyperfluid, $\rho_c$ and $p_c$ are, respectively, the canonical  (net) density and canonical pressure of the hyperfluid. In addition, the functions $\phi$, $\chi$, $\psi$, $\omega$, $\zeta$ characterize the microscopic properties of the hyperfluid which, upon using the connection field equations, act as sources of the non-Riemannian background. The tensors \eqref{metrenmomform}, \eqref{canonenmomform}, and \eqref{hypermomform}  all respect spatial isotropy and are subject to the following conservation laws\footnote{Of course these conservation laws are fairly general and hold true irrespectively of the form of the fluid one considers.}
\begin{align}
& \frac{1}{\sqrt{-g}} \hat{\nabla}_\mu \left( \sqrt{-g} {t^\mu}_\alpha \right) = \frac{1}{2} \Delta^{\lambda \mu \nu} R_{\lambda \mu \nu \alpha} + \frac{1}{2} Q_{\alpha \mu \nu} T^{\mu \nu} + 2 S_{\alpha \mu \nu} t^{\mu \nu} \,, \label{conslawshyp1} \\
& {t^\mu}_\lambda = {T^\mu}_\lambda - \frac{1}{2 \sqrt{-g}} \hat{\nabla}_\nu \left( \sqrt{-g} {\Delta_\lambda}^{\mu \nu} \right) \,. \label{conslawshyp2}
\end{align}
Eq. \eqref{conslawshyp1} follows from diffeomorphism invariance, while eq. \eqref{conslawshyp2} originates from the $\mathrm{GL}(n,\mathbb{R})$ invariance of the matter part of the action \cite{Iosifidis:2020gth} when working in the exterior calculus formalism. They will be fundamental in the cosmological study of the theory we are going to introduce. Observe that, as already anticipated, \eqref{conslawshyp2} coincides with eq. \eqref{cemt1}. Moreover, from \eqref{conslawshyp1} we can see that the canonical energy-momentum tensor naturally couples to torsion, while the metric one couples to non-metricity. Note also that one can use the latter of the above equations in order to eliminate $t^{\mu\nu}$ from the former, yielding a variant conservation law  
		\beq
	\sqrt{-g}(2 \tilde{\nabla}_{\mu}T^{\mu}_{\;\;\alpha}-\Delta^{\lambda\mu\nu}R_{\lambda\mu\nu\alpha})+\hat{\nabla}_{\mu}\hat{\nabla}_{\nu}(\sqrt{-g}\Delta_{\alpha}^{\;\;\mu\nu})+2S_{\mu\alpha}^{\;\;\;\;\lambda}\hat{\nabla}_{\nu}(\sqrt{-g}\Delta_{\lambda}^{\;\;\;\mu\nu})=0
	\eeq
	from which one can clearly see the modifications the energy-momentum tensor receives (in comparison to GR) due to the intrinsic structure of matter (i.e. hypermomentum contributions).

\section{The quadratic theory}\label{thethy}

We shall start with the most general parity preserving MAG theory quadratic in torsion and non-metricity (cf. also \cite{Obukhov:1997zd}), which depends on $11$ parameters, in the presence of a cosmological hyperfluid, that is
\begin{equation}\label{genact}
\begin{split}
S[g,\Gamma, \varphi] & =\frac{1}{2 \kappa}\int d^{n}x \sqrt{-g} \Big[ R+ b_{1}S_{\alpha\mu\nu}S^{\alpha\mu\nu} + b_{2}S_{\alpha\mu\nu}S^{\mu\nu\alpha} + b_{3}S_{\mu}S^{\mu} \\
& + a_{1}Q_{\alpha\mu\nu}Q^{\alpha\mu\nu} + a_{2}Q_{\alpha\mu\nu}Q^{\mu\nu\alpha} + 	a_{3}Q_{\mu}Q^{\mu}+ a_{4}q_{\mu}q^{\mu} + a_{5}Q_{\mu}q^{\mu} \\
& + c_{1}Q_{\alpha\mu\nu}S^{\alpha\mu\nu}+ c_{2}Q_{\mu}S^{\mu} + 	c_{3}q_{\mu}S^{\mu} \Big] +S_{\text{hyp}} \\
& =\frac{1}{2 \kappa}\int d^{n}x \sqrt{-g} \Big[ R+ \mathcal{L}_{2} \Big] +S_{\text{hyp}} \,.
\end{split}
\end{equation}
One can show that in vacuum the above theory reduces to GR. In \cite{Obukhov:1997zd} the inclusion of the homothetic curvature $\hat{R}_{\mu\nu}$ was considered to the above action, and there it was shown that in vacuum the theory is equivalent to GR plus a Proca field.\footnote{Of course this is not surprising and has to do with the form of the homothetic curvature, which is the curl (field-strength) of the non-metricity vector $Q_\mu$.} However, in the case of FLRW cosmology that we are interested in, this term would vanish identically, so there is no need to include it in our subsequent discussion.
In the following, it will be useful to introduce also the following ``superpotentials'' \cite{Iosifidis:2019jgi}:
\begin{equation}
\begin{split}
\Omega^{\alpha\mu\nu} & := a_{1}Q^{\alpha\mu\nu}+a_{2} Q^{\mu\nu\alpha}+a_{3} g^{\mu\nu}Q^{\alpha}+a_{4}g^{\alpha\mu}q^{\nu}+a_{5}g^{\alpha\mu}Q^{\nu} \,, \\
\Sigma^{\alpha\mu\nu} & := b_{1}S^{\alpha\mu\nu}+b_{2}S^{\mu\nu\alpha}+b_{3}g^{\mu\nu}S^{\alpha} \,, \\
\Pi^{\alpha\mu\nu} & := c_{1}S^{\alpha\mu\nu}+c_{2}g^{\mu\nu}S^{\alpha}+c_{3}g^{\alpha\mu}S^{\nu} \,.
\end{split}
\end{equation}
Varying \eqref{genact} with respect to the metric we derive
\beq
R_{(\mu\nu)}-\frac{R}{2}g_{\mu\nu}-\frac{\mathcal{L}_{2}}{2}g_{\mu\nu}+\frac{1}{\sqrt{-g}}\hat{\nabla}_{\alpha}\Big[ \sqrt{-g}({W^{\alpha}}_{(\mu\nu)}+{\Pi^{\alpha}}_{(\mu\nu)})\Big] +A_{(\mu\nu)}+B_{(\mu\nu)}+C_{(\mu\nu)}=\kappa T_{\mu\nu} \,, \label{metricf}
\eeq
where $\hat{\nabla}$ is defined in \eqref{hatnabla} and
\beq\label{Wtensor}
{W^{\alpha}}_{(\mu\nu)}:=2 a_{1}{Q^{\alpha}}_{\mu\nu}+2 a_{2}{Q_{(\mu\nu)}}^{\alpha}+(2 a_{3}Q^{\alpha}+a_{5}q^{\alpha})g_{\mu\nu}+(2 a_{4}q_{(\mu} + a_{5}Q_{(\mu})\delta^{\alpha}_{\nu)} \,,
\eeq
along with
\begin{equation}
\begin{split}
A_{\mu\nu} & := a_{1}(Q_{\mu\alpha\beta}{Q_{\nu}}^{\alpha\beta}-2 Q_{\alpha\beta\mu}{Q^{\alpha\beta}}_{\nu})-a_{2}Q_{\alpha\beta(\mu}{Q^{\beta\alpha}}_{\nu)} +a_{3}(Q_{\mu}Q_{\nu}-2 Q^{\alpha}Q_{\alpha\mu\nu})-a_{4}q_{\mu}q_{\nu}-a_{5}q^{\alpha}Q_{\alpha\mu\nu} \,, \\
B_{\mu\nu} & := b_{1}(2S_{\nu\alpha\beta}{S_{\mu}}^{\alpha\beta}-S_{\alpha\beta\mu}{S^{\alpha\beta}}_{\nu})-b_{2}S_{\nu\alpha\beta}{S_{\mu}}^{\beta\alpha}+b_{3}S_{\mu}S_{\nu} \,, \\
C_{\mu\nu} & := \Pi_{\mu\alpha\beta}{Q_{\nu}}^{\alpha\beta}	-( c_{1}S_{\alpha\beta\nu}{Q^{\alpha\beta}}_{\mu}+c_{2}S^{\alpha}Q_{\alpha\mu\nu}+c_{3}S^{\alpha}Q_{\mu\nu\alpha})=c_{1}({Q_{\mu}}^{\alpha\beta}S_{\nu\alpha\beta}-S_{\alpha\beta\mu}{Q^{\alpha\beta}}_{\nu})+c_{2}(S_{\mu}Q_{\nu}-S^{\alpha}Q_{\alpha\mu\nu}) \,.
\end{split}
\end{equation}
On the other hand, varying the action with respect to the general affine connection we get the field equations
\beq \label{conn2}
{P_{\lambda}}^{\mu\nu}+{\Psi_{\lambda}}^{\mu\nu}=\kappa {\Delta_{\lambda}}^{\mu\nu} \,,
\eeq
where
\begin{equation}\label{palatinidefin}
\begin{split}
{P_{\lambda}}^{\mu\nu} & := -\frac{\nabla_{\lambda}(\sqrt{-g}g^{\mu\nu})}{\sqrt{-g}}+\frac{\nabla_{\sigma}(\sqrt{-g}g^{\mu\sigma})\delta^{\nu}_{\lambda}}{\sqrt{-g}} +2(S_{\lambda}g^{\mu\nu}-S^{\mu}\delta_{\lambda}^{\nu}+g^{\mu\sigma}{S_{\sigma\lambda}}^{\nu}) \\
& = \delta_\lambda^\nu \left( q^\mu - \frac{1}{2} Q^\mu - 2 S^\mu \right) + g^{\mu \nu} \left( \frac{1}{2} Q_\lambda + 2 S_\lambda \right) - \left( {Q_\lambda}^{\mu \nu} + 2 {S_\lambda}^{\mu \nu} \right)
\end{split}
\end{equation}	
is the Palatini tensor and	
\begin{equation}
{\Psi_{\lambda}}^{\mu\nu} = {H^{\mu\nu}}_{\lambda}+\delta^{\mu}_{\lambda}k^{\nu}+\delta^{\nu}_{\lambda}h^{\mu}+g^{\mu\nu}h_{\lambda}+f^{[\mu}\delta^{\nu ]}_{\lambda} \,,
\end{equation}
with
\begin{equation}\label{usefquant}
\begin{split}
{H^{\mu\nu}}_{\lambda} & := 4 a_{1}{Q^{\nu\mu}}_{\lambda}+2 a_{2}({Q^{\mu\nu}}_{\lambda}+{Q_{\lambda}}^{\mu\nu})+2 b_{1}{S^{\mu\nu}}_{\lambda} +2 b_{2}{S_{\lambda}}^{[\mu\nu]}+c_{1}( {S^{\nu\mu}}_{\lambda}-{S_{\lambda}}^{\nu\mu}+{Q^{[\mu\nu]}}_{\lambda}) \,, \\
k_{\mu} & := 4 a_{3}Q_{\mu}+2 a_{5}q_{\mu}+2 c_{2}S_{\mu} \,, \\
h_{\mu} & := a_{5} Q_{\mu}+2 a_{4}q_{\mu}+c_{3}S_{\mu} \,, \\
f_{\mu} & := c_{2} Q_{\mu}+ c_{3}q_{\mu}+2 b_{3}S_{\mu} \,.
\end{split}
\end{equation}
Let us now elaborate a little bit on the metric field equations and derive some useful general results which we will then specialize in the case of FLRW cosmology.
To start with, we first take the trace of \eqref{metricf} to arrive at
\beq\label{trace}
\Big( 1-\frac{n}{2} \Big) R+\Big( 1-\frac{n}{2} \Big) \mathcal{L}_{2}-\frac{1}{\sqrt{-g}}\tilde{\nabla}_{\alpha}\Big[\sqrt{-g}(\Pi^{\alpha}+W^{\alpha})\Big]=\kappa T \,,
\eeq
where we recall that $\tilde{\nabla}$ denotes the Levi-Civita covariant derivative and
\begin{equation}
\begin{split}
\Pi^{\alpha} & := {\Pi^{\alpha}}_{\mu\nu}g^{\mu\nu}=(c_{1}+n c_{2}+c_{3})S^{\alpha} \,, \\
W^{\alpha}& := {W^{\alpha}}_{\mu\nu}g^{\mu\nu}=(2 a_{1}+2 n a_{3}+a_{5})Q^{\alpha}+(2 a_{2} +2 a_{4}+n a_{5})q^{\alpha} \,.
\end{split}
\end{equation}
Expanding the Ricci scalar into its Riemannian part (i.e., $\tilde{R}$) plus non-Riemannian contributions, eq. \eqref{trace} becomes
\begin{equation}\label{effEin}
\begin{split}
& \tilde{R} + \Big( a_{1}+\frac{1}{4}\Big)  Q_{\alpha\mu\nu}Q^{\alpha\mu\nu} +
\Big( a_{2}-\frac{1}{2}\Big)Q_{\alpha\mu\nu}Q^{\mu\nu\alpha} + \Big( a_{3}-\frac{1}{4}\Big)Q_{\mu}Q^{\mu}+
a_{4}q_{\mu}q^{\mu}+ \Big( a_{5}+\frac{1}{2}\Big)Q_{\mu}q^{\mu} \\
& +(b_{1}+1)S_{\alpha\mu\nu}S^{\alpha\mu\nu} + (b_{2}-2)S_{\alpha\mu\nu}S^{\mu\nu\alpha} +
(b_{3}-4)S_{\mu}S^{\mu} + (c_{1}+2)Q_{\alpha\mu\nu}S^{\alpha\mu\nu}+ (c_{2}-2)Q_{\mu}S^{\mu} +
(c_{3}+2)q_{\mu}S^{\mu} \\
& +\frac{1}{\sqrt{-g}}\tilde{\nabla}_{\mu}\Big[ \sqrt{-g}\Big( (2a_{2}+ 2 a_{4}+n a_{5}+1)q^{\mu}+(2 a_{1}+2 n a_{3}+a_{5}-1)Q^{\mu}+(c_{1}+n c_{2}+c_{3}-4)S^{\mu} \Big) \Big] =\kappa T \,.
\end{split}
\end{equation}
In addition, contracting \eqref{metricf} with $u^{\mu}u^{\nu}$ it follows that
\beq
R_{\mu\nu}u^{\mu}u^{\nu}+\frac{1}{2}(R+\mathcal{L}_{2})=\kappa T_{\mu\nu}u^{\mu}u^{\nu}+\frac{u^{\mu}u^{\nu}}{\sqrt{-g}}(\nabla_{\alpha}-2S_{\alpha})\Big[ \sqrt{-g}({\Pi^{\alpha}}_{\mu\nu}+{W^{\alpha}}_{\mu\nu})\Big] -(A_{\mu\nu}+B_{\mu\nu}+C_{\mu\nu})u^{\mu}u^{\nu} \,.	
\eeq
Then, using equation \eqref{trace} in order to eliminate the $(R+\mathcal{L}_2)$ term and expanding the covariant derivative, after some algebra we finally arrive at
\begin{equation}\label{Ruuexpr}
\begin{split}
R_{\mu\nu}u^{\mu}u^{\nu} & =\frac{\kappa}{(n-2)}	\Big[ T+(n-2)T_{\mu\nu}u^{\mu}u^{\nu}\Big] -(A_{\mu\nu}+B_{\mu\nu}+C_{\mu\nu})u^{\mu}u^{\nu} \\
& + \frac{1}{\sqrt{-g}}\partial_{\alpha}\left[ \sqrt{-g}\Big(\frac{\Pi^{\alpha}+W^{\alpha}}{(n-2)}+M^{\alpha} \Big)   \right] -2({\Pi^{\alpha}}_{(\mu\nu)}+{W^{\alpha}}_{(\mu\nu)})u^{\mu}\nabla_{\alpha}u^{\nu} \,,
\end{split}
\end{equation}
with
\begin{equation}\label{Ruu}
\begin{split}
(A_{\mu\nu}+B_{\mu\nu}+C_{\mu\nu})u^{\mu}u^{\nu} & = a_{1}u^{\mu}Q_{\mu\alpha\beta}{Q_{\nu}}^{\alpha\beta}u^{\nu}-Q_{\alpha\beta\mu}u^{\mu}(2 a_{1}Q^{\alpha\beta\nu}u_{\nu}+a_{2}Q^{\beta\alpha\nu}u_{\nu})+a_{3}(Q_{\mu}u^{\mu})^{2}-a_{4}(q_{\nu}u^{\nu})^{2} \\
& - Q_{\alpha\mu\nu}u^{\mu}u^{\nu}(2 a_{3}Q^{\alpha}+a_{5}q^{\alpha})+2 b_{1}u^{\nu}S_{\nu\alpha\beta}u_{\mu}S^{\mu\alpha\beta}-b_{1}S_{\alpha\beta\mu}u^{\mu}{S^{\alpha\beta}}_{\nu}u^{\nu} - b_{2}u^{\nu}S_{\nu\alpha\beta}u_{\mu}S^{\mu\beta\alpha} \\
& + b_{3}(S_{\mu}u^{\mu})^{2} +c_{1}u_{\nu}S^{\nu\alpha\beta}u^{\mu}Q_{\mu\alpha\beta}-c_{1}S_{\alpha\beta\mu}u^{\mu}Q^{\alpha\beta\nu}u_{\nu}+c_{2}(S_{\mu}u^{\mu})(Q_{\nu}u^{\nu})-c_{2}S^{\alpha}Q_{\alpha\mu\nu}u^{\mu}u^{\nu} \,,
\end{split}
\end{equation}
and where we have defined
\beq
M^{\alpha}:=({\Pi^{\alpha}}_{\mu\nu}+{W^{\alpha}}_{\mu\nu})u^{\mu}u^{\nu} \,.
\eeq
Note that all of the above was fairly general. In what follows we apply these considerations in cosmology. 
	
\section{Cosmology of quadratic MAG}\label{cosmquadrMAG}
	
We now wish to study the cosmology of the generalized quadratic action previously introduced. To this end, we shall consider a flat (i.e., $K=0$) $n$-dimensional\footnote{In order to keep full generality we will do our calculations for arbitrary spacetime dimension $n$ and only set $n=4$ in Subsection \ref{subcaseonlytwo}, when discussing in detail a particular sub-case of the general quadratic theory.} FLRW Universe with the usual Robertson-Walker line element as given eq. \eqref{RWle} and the homogeneous and isotropic affine connection $(\ref{connFLRW})$. The various expressions in this cosmological setup have been presented earlier in Subsection \ref{cosmtornonmetsubsec}. As already mentioned, the matter content we consider will be that of a perfect hyperfluid (cf. Subsection \ref{perfhypsubs}). Now, interestingly, due to the high symmetry of the FLRW spacetime, not all of the quadratic torsion and non-metricity invariants are independent. Indeed, as we have previously shown in Section \ref{tb} and recall here, we have the following relations:
\begin{equation}
\begin{split}
& (n-1)S_{\mu\nu\alpha}S^{\mu\nu\alpha}-(n-1)S_{\mu\nu\alpha}S^{\alpha\mu\nu}- 3 S_{\mu}S^{\mu}=0 \,, \\
& Q_{\mu}S^{\mu}-q_{\mu}S^{\mu}-(n-1)Q_{\mu\nu\alpha}S^{\mu\nu\alpha}=0 \,, \\
& (1-n) X_{1}+ (n-1)X_{2}+X_{3}-2X_{4}+X_{5}=0 \,,
\end{split}
\end{equation}
indicating that there is one relation for each of the pure torsion, pure non-metricity, and mixed scalars. As a result, using the above relations, ``only'' $8$ out of the $11$ invariants are linearly independent. We may therefore set to zero one parameter\footnote{Obviously, this is always possible since we can collect the various parameters after using the dependence relations and a renaming would amount to setting one of each equal to zero.} from each set $\{a_{I}\}$, $\{b_{J}\}$, $\{c_{J}\}$, with $I=1,2,\ldots,5$, $J=1,2,3$. We choose $\{a_{2}=0, b_{2}=0, c_{1}=0\}$, such that the generalized quadratic action boils down to
\begin{equation}\label{genact2}
\begin{split}
S[g,\Gamma, \varphi] & = \frac{1}{2 \kappa}\int d^{n}x \sqrt{-g} \Big[ R+ b_{1}S_{\alpha\mu\nu}S^{\alpha\mu\nu} + 	b_{3}S_{\mu}S^{\mu} \\
& + a_{1}Q_{\alpha\mu\nu}Q^{\alpha\mu\nu} + 	a_{3}Q_{\mu}Q^{\mu}+ a_{4}q_{\mu}q^{\mu}+ a_{5}Q_{\mu}q^{\mu} \\
& + c_{2}Q_{\mu}S^{\mu} + c_{3}q_{\mu}S^{\mu} \Big] +S_{\text{hyp}}
\end{split}
\end{equation}
and, consequently, the first of \eqref{usefquant} becomes
\begin{equation}
{H^{\mu\nu}}_{\lambda} = 4 a_{1}{Q^{\nu\mu}}_{\lambda}+2 b_{1}{S^{\mu\nu}}_{\lambda} \,.
\end{equation} 
We see that this observation has considerably simplified the analysis. The next step is to express the torsion and non-metricity variables in terms of their sources (hypermomentum variables). To this end, we start by the connection field equations \eqref{conn2} and first take its totally antisymmetric part to arrive at
\beq
2(b_{1}-1)S^{[\alpha\mu\nu]}=\kappa \Delta^{[\alpha\mu\nu]} \,,
\eeq
where we have used the fact that $P^{[\alpha\mu\nu]}=-2 S^{[\alpha\mu\nu]}$ and also $H^{[\alpha\mu\nu]}=2b_{1}S^{[\mu\nu\alpha]}=2b_{1}S^{[\alpha\mu\nu]}$. Furthermore, using the cosmological forms of the hypermomentum and Palatini tensors (cf. eqs. \eqref{hypermomform} and \eqref{palatinicosm}, respectively) we find
\beq
2(b_{1}-1)P(t)=\kappa \zeta(t) \,.
\eeq
From the latter we see that the pseudo-scalar degree of freedom completely decouples from the rest of the modes, as expected. Note that it is now essential to assume that $b_{1}\neq 1$, since otherwise $P(t)$ would be completely undetermined and $\zeta(t)$ would be forced to vanish. In addition, for the special case $b_{1}=1$ it can be easily shown that the whole action above enjoys invariance under connection transformations of the form (in four dimensions)
\beq
{\Gamma^{\lambda}}_{\mu\nu} \to {\Gamma^{\lambda}}_{\mu\nu} + {\epsilon^{\lambda}}_{\mu\nu\alpha}K^{\alpha} \,,
\eeq
being $K^a$ a generic pseudo-vector. As a result, only matter with vanishing totally antisymmetric hypermomentum part is allowed, constraining the form of matter that one can couple. We shall therefore regard the special case $b_{1}=1$ as ``unphysical'' and  assume that $b_{1}\neq 1$ for the rest of our analysis. Now, continuing we consider independent operations on the connection field equations \eqref{conn2}. First, we multiply (and contract) by $u^{\lambda}u_{\mu}u_{\nu}$, then by $u^{\lambda}h_{\mu\nu}$, subsequently in ${h^{\lambda}}_{\mu}u_{\nu}$, and finally in ${h^{\lambda}}_{\nu}u_{\mu}$, to derive the following four equations:
\begin{align}
& 2(n-1)(2a_{3}+a_{5})A(t) +(n-1)\Big( \frac{1}{2}+a_{5}+2 a_{4} \Big)B(t)-4(a_{1}+a_{3}+a_{4}+a_{5})C(t)+2(n-1)(c_{2}+c_{3})\Phi(t)=-\kappa \omega(t) \,, \\
& \Big[\frac{(n-3)}{2}+(n-1)a_{5} \Big]A(t)+\Big[2 a_{1}+(n-1)a_{4} \Big]B(t)-\Big(\frac{1}{2}+2a_{4}+a_{5}\Big) C(t)+\Big[ 2(n-2)+(n-1)c_{3}\Big] \Phi(t)=\kappa \psi(t) \,, \\
& \left[ 4 a_{1}+4(n-1)a_{3}-\frac{(n-1)}{2}c_{2}\right] A(t)+\left[ (n-1) a_{5}-\frac{(n-1)}{4}c_{3}  -\frac{1}{2} \right] B(t) +\left[ \frac{1}{2}(c_{2}+c_{3})-4 a_{3}-2 a_{5} \right]C(t) \nonumber \\
& + \Big[ 2(n-1)c_{2}-2 b_{1}-(n-1) b_{3} \Big] \Phi(t)=\kappa \phi(t) \,, \\
& (n-1) \left( -\frac{1}{2}+a_{5}+\frac{c_{2}}{2} \right) A(t)+\left[ \frac{(n-2)}{2}+2 a_{1}+(n-1) a_{4}+\frac{(n-1)}{4}c_{3} \right] B(t)-\left[ \frac{1}{2}+ 2 a_{4}+a_{5}+\frac{c_{2}+c_{3}}{2} \right]C(t) \nonumber \\
& +\Big[-2(n-2)+ 2 b_{1}+(n-1)c_{3}+(n-1)b_{3}  \Big]\Phi(t)=\kappa \chi(t) \,.
\end{align}
We may rewrite the above as
\beq\label{e1}
\alpha_{11}A+\alpha_{12}B+\alpha_{13}C+\alpha_{14}\Phi=-\kappa \omega \,,
\eeq
\beq\label{e2}
\alpha_{21}A+\alpha_{22}B+\alpha_{23}C+\alpha_{24}\Phi= \kappa \psi \,,
\eeq
\beq\label{e3}
\alpha_{31}A+\alpha_{32}B+\alpha_{33}C+\alpha_{34}\Phi= \kappa \phi \,,
\eeq
\beq\label{e4}
\alpha_{41}A+\alpha_{42}B+\alpha_{43}C+\alpha_{44}\Phi= \kappa \chi \,,
\eeq
respectively, where the relations between the $\alpha_{ij}$ coefficients and the $a$'s, $b$'s, $c$'s is obvious. For the sake of completeness, however, we include their exact relations in Appendix \ref{coeffsMAG}. We then define the matrix $M$ with elements $\alpha_{ij}$ and also the vectors $\mathcal{U}=(A,B,C,\Phi)^{T}$ and $\mathcal{W}=\kappa(-\omega,\psi,\phi,\chi)^{T}$. Hence, we can write the above equations in the matrix form
\beq
M \mathcal{U}=\mathcal{W} \,.
\eeq
Furthermore, the determinant of $M$ is non-zero \textit{in general}, since otherwise this would imply that there is some symmetry under connection transformations that would make the trace equations linearly dependent \cite{Iosifidis:2019jgi,Iosifidis:2018zwo}. Since there is no reason to demand that there is such a symmetry at play, we can safely assume that $\text{det}(M)\neq 0$ and therefore that there exists the inverse matrix $M^{-1}$. Hence, we can formally write
\beq
\mathcal{U}=M^{-1}\mathcal{W} \,.
\eeq
This equation allows to express the torsion and non-metricity variables in terms of the hypermomentum sources as
\begin{equation}
\begin{split}
A & = \kappa \left( - \lambda_{11}\omega +\lambda_{12}\psi +\lambda_{13}\phi +\lambda_{14}\chi \right) \,, \\
B & = \kappa \left( - \lambda_{21}\omega +\lambda_{22}\psi +\lambda_{23}\phi +\lambda_{24}\chi \right) \,, \\
C & = \kappa \left( - \lambda_{31}\omega +\lambda_{32}\psi +\lambda_{33}\phi +\lambda_{34}\chi \right) \,, \\
\Phi & = \kappa \left( - \lambda_{41}\omega +\lambda_{42}\psi +\lambda_{43}\phi +\lambda_{44}\chi \right) \,,
\end{split}
\end{equation}
where the $\lambda$'s are, in fact, the elements of $M^{-1}$ and are therefore given in terms of the $a$'s, $b$'s, and $c$'s by inverting $M$.
Generally, $A$, $B$, $C$, $\Phi$ depend on all four sources $\phi$, $\chi$, $\psi$, $\omega$, while the pseudo-scalar mode $P(t)$ only on $\zeta(t)$, as previously seen and actually expected. 
In the following, we will conclude the cosmological analysis of the (general) theory by deriving the associated Friedmann equations.

\subsection{Friedmann equations with torsion and non-metricity}\label{friedmanneqsthy}

Let us now derive the Friedmann equations in the presence of both torsion and non-metricity. Starting from \eqref{effEin} and considering the cosmological forms \eqref{tornonmetFLRW} of torsion and non-metricity, after some straightforward algebra we finally arrive at
\begin{equation}\label{varfr1}
\begin{split}
& 2 (n-1) \left[\frac{\ddot{a}}{a}+\frac{(n-2)}{2}\left(\frac{\dot{a}}{a}\right)^{2} \right] -(n-1)\Big[ a_{1}-\frac{(n-2)}{4}+(n-1)a_{3}\Big] A^{2}-\frac{(n-1)}{4}\Big[ 2 a_{1}+a_{2}+(n-1)a_{4}\Big] B^{2} \\ 
& -(a_{1}+a_{2}+a_{3}+a_{4}+a_{5})C^{2} -(n-1)\Big[ 2 b_{1}-b_{2}+(n-1) b_{3}-4(n-2) \Big] \Phi^{2}+ 6(b_{1}+b_{2}-1)P^{2} \delta^n_4 \\
& -(n-1)\Big[ a_{2}+\frac{(n-1)}{2}a_{5}+\frac{(n-3)}{4}\Big] A B+(n-1)(2 a_{3}+ a_{5}) AC +(n-1)\Big( a_{4}+\frac{1}{2}a_{5}+\frac{1}{4}\Big)B C \\
& -(n-1)\Big[ c_{1}+(n-1) c_{2}-2(n-2)\Big] A \Phi+\frac{(n-1)}{2}\Big[ c_{1}-(n-1) c_{3}-2(n-2)\Big] B \Phi +(n-1)(c_{2}+c_{3})C \Phi \\
& =-\Big[\dot{f}+(n-1)H f\Big]-\kappa\Big[ -\rho+(n-1)p \Big] \,,
\end{split}
\end{equation}
where we have also used \eqref{metrenmomform} and introduced
\beq
\begin{split}
f & := (2a_{2}+ 2 a_{4}+n a_{5}+1)\frac{(n-1)}{2}B-\Big[2 (a_{1}+a_{2} + n a_{3}+a_{4})+(n+1)a_{5}\Big]C \\
& +(2 a_{1} + 2 n a_{3}+a_{5}-1)(n-1)A+(c_{1}+n c_{2}+c_{3}-4)(n-1)\Phi \,.
\end{split}
\eeq
Eq. \eqref{varfr1} is a variant of the first modified Friedmann equation. Now, the acceleration equation (also known as Raychaudhuri equation, that is the second Friedmann equation) for non-Riemannian Universes has been derived in \cite{Iosifidis:2020zzp} and reads
\begin{equation}
\frac{\ddot{a}}{a}=-\frac{1}{(n-1)}R_{\mu\nu}u^{\mu}u^{\nu}+2\left( \frac{\dot{a}}{a} \right)\Phi +2\dot{\Phi} +\left( \frac{\dot{a}}{a} \right)\left(A+\frac{C}{2}\right) +\frac{\dot{A}}{2}-\frac{A^{2}}{2}-\frac{1}{2}AC -2A\Phi-2C \Phi \,.
\end{equation}
In order to find its exact form in the present case, we have to compute the term \eqref{Ruu} using the cosmological expressions of torsion and non-metricity. After some lengthy algebra, we finally find (see Appendix \ref{usefulrelFLRW} for details)
\begin{equation}\label{fr2}
\begin{split}
& \frac{\ddot{a}}{a}=-\frac{\kappa}{(n-1)(n-2)}\Big[(n-3)\rho+(n-1)p\Big] \\
& + \Big[ a_{1}+(n-1)a_{3}\Big] A^{2}-\frac{1}{4}\Big[ (2 a_{1}+a_{2})+ (n-1)a_{4} \Big] B^{2} + \frac{1}{(n-1)}( a_{1} + a_{2}+a_{3}+a_{4}+a_{5})C^{2} \\
& + \Big( a_{4}+\frac{a_{5}}{2}\Big)B C + \Big[ 2 b_{1}-b_{2}+(n-1)b_{3}\Big] \Phi^{2}+ \Big[ c_{1}+(n-1)c_{2}\Big] \Phi A \\
& - (H-Y)\Big[ \gamma_{1} B+2\gamma_{2}A+ 2 \gamma_{3}\Phi\Big]-\frac{1}{2}C \Big[ (a_{5}+2 a_{4})B+2(2 a_{3}+a_{5})A+2(c_{2}+c_{3})\Phi \Big] \\
& +(2 a_{4}+a_{5})(H-Y) C - \frac{\dot{l}}{(n-1)} - H l \\
& +2\left( \frac{\dot{a}}{a} \right)\Phi +2\dot{\Phi}
+\left( \frac{\dot{a}}{a} \right)\left(A+\frac{C}{2}\right) +\frac{\dot{A}}{2}-\frac{A^{2}}{2}-\frac{1}{2}AC 
-2A\Phi-2C \Phi \,,
\end{split}
\end{equation}
being $\gamma_1$, $\gamma_2$, $\gamma_3$, and $l$ are respectively given in \eqref{gammacoeff} and \eqref{elle} of Appendix \ref{usefulrelFLRW}, where we have also written explicitly the coefficients appearing inside $l$ (see \eqref{betacoeffs}). Recall also that $Y=2 \Phi + \frac{A}{2}$ (cf. Section \ref{tb}).
Obviously, to recover the Friedmann equations in the case of the theory with the choice $\{a_{2}=0, b_{2}=0, c_{1}=0\}$ we have previously made in a consistent way, it is sufficient to set the aforementioned coefficients to zero in the expressions above, as the latter have been derived for the general quadratic theory.
We have therefore obtained the cosmological equations. As we are left with many free parameters, in order to give more details in the following we shall focus on the case in which only the two quadratic contributions $b_3 S_\mu S^\mu$ and $a_3 Q_\mu Q^\mu$ are there, setting all other coefficients in the general quadratic model to zero. We perform this particular choice as it involves two vectors that are particularly relevant in the context of non-Riemannian geometry, namely the torsion vector $S_\mu$ and the non-metricity vector $Q_\mu$. Additionally, as we will see in what follows, these two quadratic terms not only do they simplify the analysis greatly but also they are fairly representative and give reliable conclusions to be expected also for the full quadratic theory. 

\subsection{Sub-case involving only two quadratic contributions}\label{subcaseonlytwo}

Let us now restrict ourselves to the following quadratic theory, sub-case of the general one previously discussed, in $n=4$ spacetime dimensions:
\begin{equation}\label{act3}
S[g,\Gamma, \varphi] =\frac{1}{2 \kappa}\int d^{4}x \sqrt{-g} \Big[ R+ b_{3}S_{\mu}S^{\mu} + a_{3}Q_{\mu}Q^{\mu} \Big] +S_{\text{hyp}} \,,
\end{equation}
that is we set $b_1=a_1=a_2=a_4=a_5=c_1=c_2=c_3=0$ in the general quadratic model. The metric and connection field equations can be immediately derived by implementing these constraints in the corresponding equations derived for the general theory. We will now analyze the cosmology of \eqref{act3}. From the connection field equations we get the following relations:
\begin{equation}
\begin{split}
& P = - \frac{\kappa}{2} \zeta \,, \\
& 12 a_{3} A + \frac{3}{2} B - 4 a_{3} C = -\kappa \omega \,, \\
& \frac{1}{2} A - \frac{1}{2} C + 4 \Phi = \kappa \psi \,, \\
& 12 a_{3} A -\frac{1}{2} B -4 a_{3} C - 3 b_{3} \Phi = \kappa \phi \,, \\
& -\frac{3}{2} A + B - \frac{1}{2} C + \left( -4 + 3 b_{3} \right) \Phi = \kappa \chi \,.
\end{split}
\end{equation}
The latter can be solved giving the expressions of the torsion and non-metricity variables in terms of the hypermomentum sources, namely
\begin{equation}\label{solspec}
\begin{split}
& P = - \frac{\kappa}{2} \zeta \,, \\
& A = \frac{\kappa}{[16 a_3 (3 b_3-32)-18 b_3]} \Big \lbrace (64 a_3+3 b_3+8) \omega + [-16 a_3 (3 b_3-8)-9 b_3] \psi + [64 a_3+9 b_3-24)] \phi + (128 a_3+9 b_3) \chi \Big \rbrace \,, \\
& B = \frac{\kappa}{[8 a_3 (3 b_3 -32) -9 b_3]} \Big \lbrace 2[ a_3 (64 - 12 b_3) + 3 b_3 ] \omega + 72 a_3 b_3 \psi - [8 a_3 (3 b_3-16)] \phi - 24 a_3 b_3 \chi \Big \rbrace \,, \\
& C = \frac{\kappa}{[16 a_3 (3 b_3-32)-18 b_3]} \Big \lbrace (192 a_3 + 3 b_3 -8) \omega + [27 b_3 - 48 a_3 (3 b_3-8)] \psi + 3 (64 a_3 + 3 b_3 + 8) \phi + (384 a_3 + 9 b_3) \chi \Big \rbrace \,, \\
& \Phi = \frac{\kappa}{[8 a_3 (3 b_3 -32) -9 b_3]} \Big \lbrace - (1 - 8 a_3) \omega - 48 a_3 \psi + (8 a_3 + 3) \phi + 16 a_3 \chi \Big \rbrace \,.
\end{split}
\end{equation}
Furthermore, the first modified Friedmann equation \eqref{varfr1} becomes
\begin{equation}\label{varfr1spec}
\begin{split}
& 6 \left[\frac{\ddot{a}}{a}+ \left(\frac{\dot{a}}{a}\right)^{2} \right] -3\Big(-\frac{1}{2}+3a_{3}\Big) A^{2}- a_{3} C^{2} - 3 \Big( 3 b_{3}-8 \Big) \Phi^{2} -6 P^{2} - \frac{3}{4} A B+ 6 a_{3} AC + \frac{3}{4} B C \\
& + 12 A \Phi - 6 B \Phi =-\Big(\dot{f}+3H f\Big)-\kappa\Big( -\rho+3p \Big) \,,
\end{split}
\end{equation}
where now
\beq
f = \frac{3}{2}B - 8 a_{3} C + 3 (8 a_{3}-1) A - 12 \Phi \,,
\eeq
while the second (i.e. acceleration) Friedmann equations reads
\begin{equation}\label{fr2spec}
\frac{\ddot{a}}{a} =-\frac{\kappa}{6}(\rho+ 3 p)+\frac{a_{3}}{3}\Big( 3A-C \Big)^{2}+3 b_{3}\Phi^{2}-\frac{1}{3}\Big( \dot{l}+H l \Big)+H \Big( 2\Phi+A+\frac{C}{2} \Big)-\Big( 2 \Phi+\frac{A}{2}\Big)(A+C)+ 2 \dot{\Phi} + \frac{\dot{A}}{2} \,,
\end{equation}
where now
\beq\label{ellespec}
l =2 a_{3}(3A -C) \,.
\eeq
It is worth stressing out that the second and third term in the right-hand side of eq. \eqref{fr2spec} retain a fixed sign once the values of $a_{3}$ and $b_{3}$ are given. In particular, it is interesting to note that for $a_{3}>0$ and $b_{3}>0$ both of these non-metricity and torsion contributions accelerate the expansion. Incidentally, this also means that in order for the aforementioned terms to have an accelerating effect on the expansion, both 'mass' terms for torsion and non-metricity must have the correct (positive) sign.

We shall then consider also the conservation laws for our cosmological hyperfluid model. Here let us assume that the fluid is hypermomentum preserving, or, in other words, that the metrical energy-momentum tensor coincides with the canonical one, namely
\begin{equation}\label{tT}
t^{\mu \nu} = T^{\mu \nu} \,,
\end{equation}
which implies (cf. eqs. \eqref{metrenmomform} and \eqref{canonenmomform})
\begin{equation}\label{assu1}
\rho_c = \rho \,, \quad p_c = p \,.
\end{equation}
This will be one of our assumptions when we will focus on a particular type of hypermomentum (i.e., more precisely, pure dilation hypermomentum) in the following.
Observe that the same would also apply to the case of the unconstrained hyperfluid of \cite{Obukhov:1996mg}. In fact, recalling that we are considering a homogeneous cosmological setup, we have that any antisymmetric two index object vanishes identically. In particular, this means that $t_{[\mu \nu]}\equiv0$ and, therefore, only its symmetric part survives, i.e., $t_{\mu \nu}=t_{(\mu \nu)}$. Hence we can consistently assume \eqref{tT} to hold for our model. Consequently, the conservation laws \eqref{conslawshyp1} and \eqref{conslawshyp2} become
\begin{align}
& \tilde{\nabla}_\mu {T^\mu}_\nu = \frac{1}{2} \Delta^{\lambda \mu \nu} R_{\lambda \mu \nu \alpha} \,, \label{cl1} \\
& \hat{\nabla}_\nu \left( \sqrt{-g} {\Delta_\lambda}^{\mu \nu} \right) = 0 \,, \label{cl2}
\end{align}
respectively, where  in order to  reach to the above result we have also expanded $\hat{\nabla}_\mu {T^\mu}_\nu$
into its Levi-Civita part plus non-Riemannian contributions.
Now, plugging \eqref{metrenmomform} and
\beq\label{hypermomformspec}
\Delta^{(n)}_{\alpha \mu \nu} = \phi(t) h_{\mu \alpha} u_\nu + \chi(t) h_{\nu \alpha} u_{\mu} + \psi(t) u_{\alpha} h_{\mu \nu} + \omega(t) u_\alpha u_\mu u_\nu + \epsilon_{\alpha \mu \nu \rho} u^\rho \zeta(t)
\eeq
into the above conservation laws, and recalling that in a homogeneous cosmological setup the homothetic curvature tensor $\hat{R}_{\mu \nu}$ has to be identically vanishing, the above conservation laws respectively boil down to (see also \cite{Iosifidis:2020gth})
\begin{equation}\label{c1new}
\dot{\rho} + 3 H (\rho + p) = - \frac{1}{2} u^\mu u^\nu \left( \chi R_{\mu \nu} + \psi \check{R}_{\mu \nu} \right) \,,
\end{equation}
where $\check{R}_{\mu \nu}$ is the co-Ricci tensor, and
\begin{equation}\label{c2new}
\begin{split}
& - \delta_\lambda^\mu \frac{\partial_\nu \left( \sqrt{-g}\phi u^\nu \right)}{\sqrt{-g}} - u^\mu u_\lambda \frac{\partial_\nu \left[ \sqrt{-g} \left( \phi + \chi + \psi + \omega \right) u^\nu \right]}{\sqrt{-g}} + \left[ \left( 2 S_\lambda + \frac{Q_\lambda}{2} \right) u^\mu - \nabla_\lambda u^\mu \right] \chi \\
& + \left[ \left( 2 S^\mu + Q^\mu - q^\mu \right) u_\lambda - g^{\mu \nu} \nabla_\nu u_\lambda \right] \psi + u^\mu u_\lambda \left( \dot{\chi} + \dot{\psi} \right) - \left( \phi + \chi + \psi + \omega \right) \left( \dot{u}^\mu u_\lambda + u^\mu \dot{u}_\lambda \right) = 0 \,.
\end{split}
\end{equation}
 The former of the above equations is the modified perfect fluid continuity equation which receives corrections from the hypermomentum, while the latter provides the evolution laws for the hypermomentum currents (i.e., non-Riemannian degrees of freedom).
Therefore, we are left with eqs. \eqref{solspec}, \eqref{varfr1spec}, \eqref{fr2spec}, \eqref{c1new}, and \eqref{c2new}, ruling the cosmology of the theory.
In the following, we will complete the analysis by focusing on the cases in which specific parts of the hypermomentum tensor are switched on in order to highlight the role of its irreducible parts and their contribution separately. To be more specific, the hypermomentum tensor can be decomposed as follows \cite{Iosifidis:2020upr} (recall that we are considering four spacetime dimensions):
\begin{equation}
\Delta_{\alpha \mu \nu} = \Delta_{[\alpha \mu] \nu} + \frac{1}{4} g_{\alpha \mu} D_\nu + \breve{\Delta}_{\alpha \mu \nu} \,,
\end{equation}
where the first term on the right-hand side represents the spin part, $D^\nu := {\Delta_\mu}^{\mu \nu}$ is the dilation, and $\breve{\Delta}_{\alpha \mu \nu}$ the shear (traceless symmetric part of the hypermomentum tensor). Then, given the most general form \eqref{hypermomform} of hypermomentum compatible with the cosmological principle (i.e. respecting both isotropy and homogeneity), the spin, dilation and shear parts read
\begin{align}
& \Delta_{[\alpha \mu ] \nu} = \left( \psi - \chi \right) u_{[\alpha} h_{\mu]\nu} + \epsilon_{\alpha \mu \nu \rho} u^\rho \zeta \,, \label{hspin} \\
& D_\nu := \Delta_{\alpha \mu \nu} g^{\alpha \mu} = \left( 3 \phi - \omega \right) u_\nu \,, \label{hdilation} \\
& \breve{\Delta}_{\alpha \mu \nu} = \Delta_{(\alpha \mu )\nu} - \frac{1}{4} g_{\alpha \mu} D_\nu = \frac{\left( \phi + \omega \right)}{4} \left( h_{\alpha \mu} + 3 u_\alpha u_\mu \right) u_\nu + \left( \psi + \chi \right) u_{(\mu} h_{\alpha ) \nu} \,, \label{hshear}
\end{align}
respectively.
Besides, for the case of pure dilation we are going to discuss in the following, we will assume that the perfect hyperfluid variables are related through a barotropic equation of state of the usual type, namely
\begin{equation}\label{barotropiceos}
p= w \rho \,,
\end{equation}
where $w$ is the associated barotropic index.
Substituting this into the acceleration equation
 \eqref{fr2spec} we may then use the latter and eliminate the double derivative term ($\ddot{a}$) from \eqref{varfr1spec}, in such a way so as to obtain the final form of the first modified Friedmann equation for the model at hand, under the above  assumptions , which reads
\begin{equation}\label{fr1final}
\begin{split}
H^2 & = \frac{1}{2} \left(1-6a_3 \right)A^2 - \frac{1}{6} a_3 C^2 + \left(-4 - \frac{3}{2} b_3 \right) \Phi^2 + P^2 + \frac{1}{8} A B + \left(\frac{1}{2} + a_3 \right) A C - \frac{1}{8} B C \\
& + 2 A \Phi + B \Phi + 2 C \Phi - 2 a_3 \dot{A} - \frac{1}{4} \dot{B} + \frac{2}{3} a_3 \dot{C} + H \left[ \left(\frac{1}{2} - 6 a_3 \right) A - \frac{3}{4} B + \left( 2 a_3 - \frac{1}{2} \right) C + 4 \Phi \right]  + \frac{\kappa}{3} \rho \,,
\end{split}
\end{equation}
where we recall that
\begin{equation}
H:=\frac{\dot{a}}{a}
\end{equation}
is the Hubble parameter. Furthermore, using \eqref{barotropiceos} into \eqref{c1new}, the latter becomes
\begin{equation}\label{c1final}
\dot{\rho} + 3 H (1+w) \rho = - \frac{1}{2} u^\mu u^\nu \left( \chi R_{\mu \nu} + \psi \check{R}_{\mu \nu} \right) \,.
\end{equation}
To recap, the two Friedmann equations   eqs. \eqref{solspec},  \eqref{fr1final} supplemented with the conservations laws \eqref{c2new} and \eqref{c1final}, close the system of equations and we can now concentrate on the cosmological effects of dilation and spin.\footnote{We leave the case of pure shear hypermomentum to future work. The role of the latter but in the case where only the Einstein-Hilbert term was present in the gravitational action was invastigated in \cite{Iosifidis:2020upr}.}

\subsubsection{Purely dilation hypermomentum}

Let us now consider the case in which the hypermomentum tensors encodes only dilation, namely $\Delta_{[\alpha \mu ] \nu}=0$ and $\breve{\Delta}_{\alpha \mu \nu}=0$.
Specifically, this implies
\begin{equation}\label{conddil}
\zeta = 0 \,, \quad \psi = 0 \,, \quad \chi = 0 \,, \quad \omega = - \phi \,,
\end{equation}
which give
\begin{equation}\label{dil}
\Delta_{\alpha \mu \nu} = \frac{1}{4} g_{\alpha \mu} D_\nu = g_{\alpha \mu} \phi u_\nu \,.
\end{equation}
Therefore, we are left just with one independent, non-vanishing source, that is $\phi$.
As a result, from \eqref{solspec} now we get
\begin{equation}\label{solspecdil}
\begin{split}
& P = 0 \,, \\
& A = \kappa \left[ \frac{(16 - 3 b_3)}{8 a_3 (32 - 3 b_3) + 9 b_3} \right] \phi \,, \\
& B = \kappa \left[ \frac{6 b_3}{8 a_3 (32 - 3 b_3) + 9 b_3} \right] \phi \,, \\
& C = - \kappa \left[ \frac{(16 + 3b_3)}{8 a_3 (32 - 3 b_3) + 9 b_3} \right] \phi \,, \\
& \Phi = - \kappa \left[ \frac{4}{8 a_3 (32 - 3 b_3) + 9 b_3} \right] \phi \,.
\end{split}
\end{equation}
As we can see, the pseudo-scalar mode $P$ vanishes, while the other ones are given in terms of $\phi$. It is interesting to note that when the ``mass'' term for the torsion vector is absent (i.e., $b_{3}=0$) it follows that $B=0$ and $A+C=0$, which is exactly the case of a Weyl non-metricity. In other words, for $b_{3}=0$, interestingly enough, the geometry is fixed to be that of Weyl-Cartan type. Note also that there exist specific values for $b_{3}$ for which one can obtain a vanishing part for all (but one each time) of the three non-metricity functions, while there is no parameter value for $b_{3}$ (or even $a_{3}$) that yields a vanishing torsion function $\Phi$.
Going back to our analysis, the conservation laws \eqref{c2new} and \eqref{c1final} respectively reduce to
\begin{align}
& \partial_\nu \left( \sqrt{-g} \phi u^\nu \right) = 0 \,, \label{cl2dil0} \\
& \dot{\rho} + 3 H (1+w) \rho = 0 \,. \label{cl1dil}
\end{align}
In particular, the latter provides the evolution of $\rho$, while from the former we get
\begin{equation}\label{cl2dil}
\dot{\phi} + 3 H \phi = 0 \,.
\end{equation}
Note that in this case the density decouples and evolves as in the usual perfect fluid continuity equation, while also for $\phi$ we observe an analogous evolution which, in particular, mimics that of dust. 
Moreover, plugging \eqref{solspecdil} into \eqref{fr1final} we are left with
\begin{equation}\label{freqdil0}
H^2 = - \frac{\kappa}{6} \dot{\phi} - \frac{\kappa}{2} H \phi - \frac{\left[ 3072 + 8 a_3 \left(32-3 b_3 \right)^2 - 288 b_3 - 81 b_3^2 \right]}{12\left[9 b_3 - 8 a_3 \left(3 b_3 - 32 \right)\right]^2} \kappa^2 \phi^2 + \frac{\kappa}{3} \rho \,.
\end{equation}
where, of course, we have $9 b_3 - 8 a_3 \left(3 b_3 - 32 \right) \neq 0$. Furthermore, using the above continuity equation for $\phi$ the latter simplifies to
\begin{equation}\label{freqdil}
H^2 = - \frac{\left[ 3072 + 8 a_3 \left(32-3 b_3 \right)^2 - 288 b_3 - 81 b_3^2 \right]}{12\left[9 b_3 - 8 a_3 \left(3 b_3 - 32 \right)\right]^2} \kappa^2 \phi^2 + \frac{\kappa}{3} \rho \,.
\end{equation}
Observe that for $\phi=0$ or $a_3 = \frac{3 (-1024 + 96 b_3 + 27 b_3^2)}{8 (1024 - 192 b_3 + 9 b_3^2)}$ one would recover the usual (Riemannian) cosmological equation in the flat (that is, $K=0$) case.\footnote{On the other hand, considering a non-flat cosmological background, namely introducing a curvature parameter $K\neq0$, one would get
\beq
H^{2}= - \frac{K}{a^2} - \frac{\left[ 3072 + 8 a_3 \left(32-3 b_3 \right)^2 - 288 b_3 - 81 b_3^2 \right]}{12\left[9 b_3 - 8 a_3 \left(3 b_3 - 32 \right)\right]^2} \kappa^2 \phi^2 + \frac{\kappa}{3} \rho \,, \nonumber
\eeq 
where we can see that the usual term induced by the presence of a non-null curvature parameter $K$ appears on the right-hand side.}
Concluding, let us also report the final form of the acceleration equation, that is
\begin{equation}\label{fr2dilfin}
\frac{\ddot{a}}{a} = \frac{\left[4 a_3 \left(32 - 3 b_3 \right)^2 + 9 \left( 16 - 3 b_3 \right) b_3 \right]}{3 \left[9 b_3 - 8 a_3 \left(3 b_3 - 32 \right)\right]^2} \kappa^2 \phi^2 - \frac{\kappa}{6} (1+3w) \rho \,.
\end{equation}
Again, note that for $\phi=0$ or $a_3 = \frac{9 \left( -16 b_3 + 3 b_3^2 \right)}{4 (1024 - 192 b_3 + 9 b_3^2)}$ one would recover the usual (purely Riemannian) acceleration equation.
We are now in a position to provide solutions.

\vspace{0.2cm}

\paragraph{Purely dilation hypermomentum solutions.}

Let us first of all collect all relevant cosmological equations obtained till now in the case of pure dilation. They are
\begin{align}
& H^2 = \frac{\kappa}{3} \rho + \mathcal{B} \frac{\kappa^2 \phi^2}{4} \,, \label{rele1} \\
& \frac{\ddot{a}}{a} = - \frac{\kappa}{6} (1+3w) \rho - \mathcal{C} \frac{\kappa^2 \phi^2}{2} \,, \label{rele2} \\
& \dot{\rho} + 3 H (1+w) \rho = 0 \,, \label{rele3} \\
& \dot{\phi} + 3 H \phi = 0 \,, \label{rele4}
\end{align}
where, for simplicity, we have introduced
\begin{align}
\mathcal{B} & := - \frac{\left[ 3072 + 8 a_3 \left(32-3 b_3 \right)^2 - 288 b_3 - 81 b_3^2 \right]}{3\left[9 b_3 - 8 a_3 \left(3 b_3 - 32 \right)\right]^2} \,, \label{mathcalB} \\
\mathcal{C} & := - \frac{\left[4 a_3 \left(32 - 3 b_3 \right)^2 + 9 \left( 16 - 3 b_3 \right) b_3 \right]}{3 \left[9 b_3 - 8 a_3 \left(3 b_3 - 32 \right)\right]^2} \,, \label{mathcalC}
\end{align}
with, of course, $9 b_3 - 8 a_3 \left(3 b_3 - 32 \right) \neq 0$.
Some comments are in order. First, notice that in eqs. \eqref{rele1} and \eqref{rele2}, in the end, there are no linear terms in $H$ and these equations have a more clear physical interpretation.
In fact, from the aforementioned equations we can see that $\phi^2$ acts as an additional perfect fluid component with some peculiar associated density and pressure which can just have a net effect that accelerates the expansion. The two Friedmann equations are then supplemented with the conservation laws \eqref{rele3} and \eqref{rele4}, and we have a closed system
of equations.
Besides, note from \eqref{rele2} that the effect of dilation hypermomentum on the cosmological expansion depends crucially on the sign of $\mathcal{C}$. More specifically, for $\mathcal{C}>0$ the dilation
hypermomentum slows down expansion, while for $\mathcal{C}<0$ it accelerates the latter and for $\mathcal{C}=0$ it has no effect on the acceleration equation. We shall therefore refer to these cases as repulsive
($\mathcal{C}>0$), attractive ($\mathcal{C}<0$), and static ($\mathcal{C}=0$) dilation, respectively (following the same lines of \cite{Iosifidis:2020upr}, where, however, shear hypermomentum was considered). Secondly, from \eqref{rele1} it follows that if $\mathcal{B}>0$ then dilation enhances the total density (and we have an effective amplification of the latter), while for $\mathcal{B}<0$ it decreases the total density and for $\mathcal{B}=0$ it does not affect the equation. This is also supported by the fact that only $\phi^2$ terms appear in the two Friedmann equations, meaning that an identification $\rho \propto \phi^2$ is possible.
Finally, differentiating \eqref{rele1} and employing both \eqref{rele3} and \eqref{rele4}, comparing the result with \eqref{rele2}, one obtains
\begin{equation}
\mathcal{C} = \mathcal{B} \,,
\end{equation}
which, in turn, implies
\begin{equation}
a_3 = \frac{3 \left[-512 + 9 b_3 (8 + b_3)\right]}{2 (32 - 3 b_3)^2} \,.
\end{equation}
Hence, we are left with a single independent parameter, $b_3$,\footnote{The relation between $a_3$ and $b_3$ we get as explained above is a peculiarity of the quadratic model we are considering under the assumptions \eqref{assu1} and \eqref{barotropiceos} in the case of pure dilation hypermomentum.} and the final system of equations reads as follows:
\begin{align}
& H^2 = \frac{\kappa}{3} \rho + \mathcal{B} \frac{\kappa^2 \phi^2}{4} \,, \label{rele1fin} \\
& \frac{\ddot{a}}{a} = - \frac{\kappa}{6} (1+3w) \rho - \mathcal{B} \frac{\kappa^2 \phi^2}{2} \,, \label{rele2fin} \\
& \dot{\rho} + 3 H (1+w) \rho = 0 \,, \label{rele3fin} \\
& \dot{\phi} + 3 H \phi = 0 \,, \label{rele4fin}
\end{align}
with
\begin{equation}
\mathcal{B} = - \frac{(32 - 3 b_3)^2 \left[ -1024 + 3 b_3 (64 + 3 b_3) \right]}{9 \left[ 2048 - 3 b_3 (128 + 9 b_3) \right]^2} \,,
\end{equation}
along with, of course, $2048 - 3 b_3 (128 + 9 b_3) \neq 0$.
We can finally recognize the following distinct cases: repulsive
($\mathcal{B}>0$, for which dilation also enhances the total density), attractive ($\mathcal{B}<0$, for which dilation also reduces the total density), and static ($\mathcal{B}=0$, for which dilation does not affect eq. \eqref{rele1fin} and \eqref{rele2fin}) dilation, respectively.

Now, as it is commonly accepted that non-Riemannian effects may have played a significant role in the early stages of the Universe, for early times one can assume dilation to be dominating with respect to the perfect fluid characteristics. Hence, in this regime, we can safely ignore the latter, namely $\rho$ and $p$, and focus only on the cosmological effects of dilation. 
By doing so, we are left with
\begin{align}
& H^2 = \mathcal{B} \frac{\kappa^2 \phi^2}{4} \,, \label{sd1} \\
& \frac{\ddot{a}}{a} = - \mathcal{B} \frac{\kappa^2 \phi^2}{2} \,, \label{sd2} 
\end{align}
while eq. \eqref{rele4fin} can be immediately integrated to get
\begin{equation}
\phi = \phi_0 \left( \frac{a_0}{a} \right)^3 \,,
\end{equation}
given that for some fixed time $t = t_0$ we have $a(t_0) = a_0$ and $\phi(t_0)=\phi_0$. On the other hand, from \eqref{sd1} we can see that we must have $\mathcal{B}>0$ (i.e., deceleration, from \eqref{sd2}). Besides, solving \eqref{sd1} for $H$, we find
\begin{equation}
H = \mathcal{B}_1 \phi \,, \quad \mathcal{B}_1 := \pm \frac{\kappa \sqrt{\mathcal{B}}}{2} \,.
\end{equation}
Then, plugging the above expression for $\phi$ into this last equation, after integration we obtain
\begin{equation}
a(t)= a_0 \left[ \mathcal{B}_1 \phi_0 \left( t-t_0 \right) + 1 \right]^{\frac{1}{3}} = a_0 \left[ \pm \frac{\kappa \sqrt{\mathcal{B}} \phi_0}{2} \left( t-t_0 \right) + 1 \right]^{\frac{1}{3}} \,.
\end{equation}
The latter describes the cosmological evolution for our model in the case of a dilation dominated Universe. Quite remarkably, we can notice that the effect of dilation is identical to that of stiff matter. To conclude the analysis regarding purely dilation hypermomentum, let us also mention that, from the two modified Friedmann equations, exploiting the definition of the deceleration parameter $q$, one easily finds
\begin{equation}
q := - \frac{\ddot{a}a}{\dot{a}^2} = 2 > 0 \,,
\end{equation}
meaning, in accordance with the fact that here we must have $\mathcal{B}>0$, that considering our model in the case of a dilation dominated (early) Universe, one finds out that the latter is subject to a cosmic deceleration.

\subsubsection{Purely spin hypermomentum}

It is also interesting to study the case in which the hypermomentum is only of pure spin type. In this instance we have that $\Delta_{\alpha\mu\nu}=\Delta_{[\alpha\mu]\nu}$ and accordingly 
 $D_\nu = 0$ as well as $\breve{\Delta}_{\alpha \mu \nu}=0$. The latter imply
\begin{equation}
\phi = 0 \,, \quad \omega = 0 \,, \quad \chi = - \psi \,,
\end{equation}
along with
\begin{equation}
\Delta_{\alpha \mu \nu} = \Delta_{[\alpha \mu ] \nu} = 2 \psi u_{[\alpha} h_{\mu]\nu} + \epsilon_{\alpha \mu \nu \rho} u^\rho \zeta \,,
\end{equation}
and as a result the non-metricity and torsion cosmological functions become
\begin{align}
& P = - \frac{\kappa}{2} \zeta \,, \label{PP} \\ 
& A = -\frac{3 b_{3}}{b_{0}}(3+8 a_{3})\kappa \psi  \,, \\
& B = \frac{ 96 a_{3} b_{3}}{b_{0}}\kappa \psi \label{bb} \,, \\
& C = \frac{ b_{3}}{b_{0}}9\left(1 - 8 a_{3}\right)\kappa \psi \,, \\
& \Phi = -\frac{64 a_{3}}{b_{0}}\kappa \psi \,, \label{Phi}
\end{align}
where we have set
\beq
b_{0}:=\frac{1}{8 a_{3}(3 b_{3}-32)- 9 b_{3}} \,.
\eeq
From the above forms we can infer some quite interesting characteristics.  In particular, we see that in order to have non-vanishing non-metricity variables it must hold that $b_{3}\neq 0$, that is the $S_{\mu}S^{\mu}$ term must be present. On the other hand, in order for the torsion function $\Phi$ not to vanish we must have that $a_{3}\neq 0$, namely the $S_{\mu}S^{\mu}$ term must be there\footnote{This is a very remarkable   duality and of course highly non-trivial or to be excepted by any speculation. }. From the above we also observe a highly non-trivial consequence: the spin part can source non-metricity (along with torsion). Let us note that for the spin part to source these non-Riemannian degrees of freedom the inclusion of the quadratic torsion and non-metricity invariants in the Lagrangian is absolutely essential.

Here, let us continue our analysis without imposing the assumptions \eqref{assu1} and \eqref{barotropiceos}, as releasing the latter, in this case, leads to interesting results. From the conservation law \eqref{conslawshyp1} we get the two relations,
\beq
p_{c}-p=\frac{1}{4}B \psi \,, \label{p}
\eeq
\beq
\rho_{c}-\rho=\frac{3}{4}B \psi \,, \label{rho}
\eeq
which also imply that\footnote{Note that this seems to be a ``radiation-like'' equation of state $\hat{p}=\hat{\rho}/3$ for the modified pressure $\hat{p}=(p_{c}-p)$ and density $\hat{\rho}=(\rho_{c}-\rho)$.}
\beq
\rho_{c}-\rho=3(p_{c}-p) \,.
\eeq
We notice, from the above equations, that it is quite crucial in this case to consider a generalized non-preserving hypermomentum, namely $\rho_{c}\neq \rho$ and $p_{c}\neq p$ need to hold true. In the case of a hypermomentum preserving hyperfluid ($\rho_{c}=\rho$, $p_{c}=p$) from the above it immediately follows that either $\psi=0$ or $B=0$. For the former possibility ($\psi=0$), the full hypermomentum would vanish, trivializing therefore the whole analysis. Also, in the latter case, given \eqref{bb}, the constraint $B=0$ would then demand either $a_{3}$ or $b_{3}$ to vanish, producing, therefore, an inconsistent theory. 
With the above in mind, we can then combine eqs. \eqref{p} and \eqref{rho} with \eqref{bb} to obtain the net density
\beq
\rho_{c}=\rho+\frac{72 a_{3} b_{3}}{b_{0}}\kappa \psi^{2} \label{rc}
\eeq
as well as the net pressure
\beq
p_{c}=p+\frac{24 a_{3} b_{3}}{b_{0}}\kappa \psi^{2} \,. \label{pc}
\eeq
From these we see that it is important to include both quadratic terms in torsion and non-metricity (i.e., $a_{3}\neq 0$ and $b_{3}\neq0$) at the same time in order to observe modifications to the total pressure and density due to the hypermomentum degrees of freedom. In addition, as it is obvious from the above exposure, depending on the sign of the ratio $a_{3}b_{3}/b_{0}$ the hypermomentum contribution can either enhance or decrease the observed density and pressure. It is interesting to note that the hypermomentum contribution to the density might as well be  negative since as long as it does not exceed $\rho$  the net density will always be positive.\footnote{This situation shares a certain similarity with the negative temperature system in thermodynamics. It is known that in given thermodynamical systems some property of the system can possess negative temperature but the net temperature will always be positive.} However, in the very early Universe there might exist an exotic hyperfluid displaying a negative total density.

Coming back to our cosmological analysis, from the modified continuity equation \eqref{conslawshyp2}, after some algebra and simplifications, it follows that
\beq
\dot{\rho}_{c}+3 H (\rho_{c}+p_{c})=-\frac{3}{2}\mu_{1} \kappa \psi \Big( \dot{\psi} +H \psi \Big)-3 \psi \frac{\ddot{a}}{a} \,,
\eeq
where
\beq
\mu_{1}:=\frac{28 a_{3}+48 a_{3} b_{3}+\frac{9}{2}b_{3}}{b_{0}} \,.
\eeq
Now, during a hypermomentum dominated era (very early Universe), the main contributions in \eqref{rc} and \eqref{pc} would be the ones $\propto \psi^{2}$. In other words, the classical perfect fluid contributions $\rho$ and $p$ can be ignored. The same considerations apply to the acceleration equation \eqref{fr2spec}, which can  therefore be written in terms of $\psi$ as follows
\beq\label{daaspin}
\frac{\ddot{a}}{a} = \frac{\kappa}{2 b_0^2} \left( \dot{\psi} + H \psi - 2304 \kappa a_3^2 b_3^2 \psi^2 \right) \,.
\eeq
From the right-hand side of the above equation, we note that the last term, keeping a fixed negative sign, always decelerates the expansion. However, there is also the presence of the first and second terms which can radically change the net result. As we will show shortly in the following, the actual effect of the spin part of hypermomentum on the acceleration depends on the parameters of the quadratic theory. 
Having obtained the acceleration equation above we can then use it in order to eliminate the double derivative term $\frac{\ddot{a}}{a}$ from the continuity equation, which results in
\beq
\dot{\psi}+ \left( 1+ \mu_{2}\right) H\psi+\mu_{3}\kappa \psi^{2}=0 \,, \label{conlawpsi}
\eeq
where we have also introduced
\begin{equation}
\begin{split}
\mu_2 & := 96 a_3 b_3 \nu_1 \,, \\
\mu_3 & := - \frac{2304}{b_0} a_3^2 b_3^2 \nu_1 \,,
\end{split}
\end{equation}
with
\begin{equation}
\nu_1 := \frac{1}{[-9 b_3 + 8 a_3 (-32 + 15 b_3) + b_0 \mu_1]} \,.
\end{equation}
On the other hand, the first modified Friedmann equation \eqref{varfr1spec} now becomes
\begin{equation}
H^2 + \frac{\ddot{a}}{a} = \frac{\kappa}{4 b_0} \Big \lbrace \frac{\kappa}{b_0} \left[ -9 b_3 + 8 a_3 (32 + 3 b_3) \right] \left[ 9 b_3 + 8 a_3 (-32 + 9 b_3) \right] \psi^2 + \frac{2}{b_0} \left( \dot{\psi} + 3 H \psi \right) \Big \rbrace + \frac{\kappa^2}{4} \zeta^2 \,.
\end{equation}
Finally, plugging \eqref{daaspin} into the above equation in order to eliminate $\frac{\ddot{a}}{a}$, we are left with 
\begin{equation}
H^2 = \frac{\kappa}{2 b_0^2} \left[ \pi_1 \dot{\psi} + (2+\pi_1) H \psi + \frac{\kappa}{2} \pi_2 \psi^2 \right] + \frac{\kappa^2}{4} \zeta^2 \,,
\end{equation}
where
\begin{equation}
\begin{split}
\pi_1 & := 1 + 9 b_0 b_3 - 8 a_3 b_0 (-32 + 3 b_3) \,, \\
\pi_2 & := -81 b_3^2 - 144 a_3 b_3 (-32 + 3 b_3) + 64 a_3^2 (-1024 + 192 b_3 + 99 b_3^2) \,.  
\end{split}
\end{equation}
Now, even though at first sight the system might not look so manageable, as we will show with some simple observations we can get exact exact solutions quite easily. 

\vspace{0.2cm}

\paragraph{Purely spin hypermomentum solutions.}

To begin with, we note that the remaining spin part $\zeta$ will be related to $\psi$ with a barotropic equation of state of the form $\zeta=w_{\zeta}\psi$, where $w_{\zeta}$ is the barotropic index related to this spin part (cf. also Appendix B of \cite{Iosifidis:2020gth}). Next we may use \eqref{conlawpsi} in order to eliminate the derivative term $\dot{\psi}$ from the modified first Friedmann equation above, ending up with 
\beq\label{quadreq}
H^{2}-\lambda_{1}H \psi-\lambda_{2}\psi^{2}=0 \,,
\eeq
where
\beq
\lambda_{1}:=\frac{\kappa}{2 b_0^2} \Big( 2 - \mu_2 \pi_1 \Big) \,, \quad \lambda_{2}:=\frac{\kappa^{2}}{2 b_0^2}\left( - \pi_1 \mu_{3} + \frac{\pi_{2}+b_{0}^{2}w_{\zeta}^{2}}{2}\right) \,.
\eeq
Then, we observe that \eqref{quadreq} is a simple quadratic equation, which, considering $H$ as the unknown variable, admits the solutions (for $\lambda_{1}^{2}+4 \lambda_{2}>0$)
\beq
H=\lambda_{0}\psi \,, \quad \lambda_{0}=\frac{\lambda_{1}\pm \sqrt{\lambda_{1}^{2}+4 \lambda_{2}}}{2} \,. \label{Hpsi}
\eeq 
Substituting this back into the continuity equation for $\psi$ it follows that
\beq
\dot{\psi}=-\mu_{0}\psi^{2} \,, \quad \mu_{0}=\lambda_{0}(1+\mu_{2})+\kappa \mu_{3} \,,
\eeq
which trivially integrates to
\beq
\psi(t)=\frac{1}{c_{1}+\mu_{0}t} \,,
\eeq
where $c_1$ is an arbitrary integration constant.
Finally, substituting this form for $\psi$ back into \eqref{Hpsi} and integrating, we find the following expression for the scale factor:
\beq
a(t)=c_{2}(c_{1}+\mu_{0}t)^{\frac{\lambda_{0}}{\mu_{0}}} \,,
\eeq
where $c_{1}$ and $c_{2}$ are integration constants that can be expressed in terms of  the values  $a_{0}$ and $\psi_{0}$ for some fixed time $t_{0}$.
Therefore, we have established exact solutions for both the scale factor and the spin hypermomentum variable $\psi$. We see that we get power-law solutions for the scale factor and inverse time evolution for the spin part. As a result, $\psi$ diminishes with the passing of time, while the scale factor goes like
\begin{equation}\label{scfactspin}
a \propto t^{\frac{\lambda_{0}}{\mu_{0}}} \,.
\end{equation}
It is also interesting to study the two limits $\mu_{0}\rightarrow 0$ and $\mu_{0}\rightarrow \infty$. In the former case the spin concentration ($\psi$) becomes constant and subsequently we get de Sitter-like expansion for the scale factor $a \propto e^{H_{0}t}$. Hence we see that a constant spin distribution produces an exponential expansion. In the latter case (i.e., $\mu_{0}\rightarrow \infty$) $\psi$ essentially vanishes, resulting also in $H=0$ and yielding, therefore, a static Universe. These cover the two extreme cases and for the rest in between we have the nice power-law solutions we derived above.

\section{Conclusions}\label{concl}

We have investigated cosmological aspects of the most general parity preserving MAG theory quadratic in torsion and non-metricity in the presence of a cosmological hyperfluid, deriving the modified Friedmann equations in a FLRW background. The theory exhibits, in principle, $11$ parameters coming as coefficients of the $11$ quadratic torsion and non-metricity scalars supplementing the Einstein-Hilbert term in the gravitational action. For this general quadratic MAG theory we have subsequently derived the most general form of the modified Friedmann equations, along with the associated conservation laws for the sources, in the presence of a cosmological hyperfluid \cite{Iosifidis:2020gth,Iosifidis:2021nra}. Then, in order to get a grasp of the effect of the quadratic terms and also simplify the analysis, we have focused on a particular but fairly representative sub-case involving only the two quadratic contributions with the parameters $a_3$ and $b_3$, namely $a_3 Q_\mu Q^\mu$ and $b_3 S_\mu S^\mu$, which are respectively quadratic in the non-metricity vector $Q_\mu$ and in the torsion vector $S_\mu$. For this case we have studied the modified Friedmann equations along with the conservation laws of the perfect cosmological hyperfluid and we have been able to provide exact solutions for particular types of hypermomentum, that of purely dilation and purely spin matter. It is worth mentioning that for the pure dilation case if the torsion squared term is absent (that is for $b_{3}=0$), quite remarkably, the geometry is fixed to be that of Weyl-Cartan type. On the other hand, for the pure spin case, interestingly, the presence of $b_{3}\neq 0$ is essential to guarantee non-vanishing non-metricity, while non-vanishing $a_{3}$ is required in order for the torsion not to be zero. This is a quite compelling result showing a dualistic relation between torsion and non-metricity, that is in order to have non-vanishing non-metricity the pure torsion term must be present and the torsion is non-vanishing if the pure non-metricity term is there, as seen from eqs. \eqref{PP}-\eqref{Phi}. 

Now, coming back to the hypermomentum parts, regarding pure dilation we have found that, for early times in which non-Riemannian effects are dominant, dilation causes a cosmic deceleration and its effect is indistinguishable from that of stiff matter.\footnote{The same result was also reported in the analysis of \cite{Obukhov:1997zd}.} A very similar result was found for the purely shear counterpart of hypermomentum that was studied in \cite{Iosifidis:2020upr}. With these two results, one could go as far as to say that the rather unconventional character of ``stiff'' matter component acquires a clear physical meaning as a specific part of the hypermomentum contribution (either shear or dilation) of the hyperfluid.    
On the other hand, concerning purely spin hypermomentum, we have obtained power-law solutions for the scale factor and inverse time evolution for the spin part $\psi$. In particular, $\psi$ diminishes with the passing of time, while the scale factor behavior is given by eq. \eqref{scfactspin}. Here, two particular limits emerge: one in which we have an exponential expansion for the scale factor induced by a constant spin distribution and the other in which the result is a static Universe ($H=0$).
We have therefore proven that the presence of torsion and non-metricity driven by dilation and spin hypermomentum types in the context of quadratic MAG can have highly non-trivial effects on the cosmological evolution. 

It would be interesting to study also the purely shear hypermomentum in this context. Another possible future development of the present work could consist in performing the cosmological analysis by including also parity odd quadratic terms into the gravitational part of the action which also include the so-called Hojman \cite{Hojman:1980kv} (sometimes also referred to Holst \cite{Holst:1995pc}) parity violating term, whose contribution in the framework of MAG has been recently presented and studied to some extent in \cite{Iosifidis:2020dck}. Finally, it would be of much interest to investigate the possible changes implied by the inclusion of mixed torsion--non-metricity terms such as $S_{\mu}Q^{\mu}$ on top of the pure torsion and pure non-metricity terms we have considered here.

\vspace{0.5cm}

\begin{center}
\textbf{Acknowledgments}
\end{center}

D.I. acknowledges: This research is co-financed by Greece and the European Union (European Social Fund - ESF) through the Operational Programme `Human Resources Development, Education and Lifelong Learning' in the context of the project ``Reinforcement of Postdoctoral Researchers - 2nd Cycle'' (MIS-5033021), implemented by the State Scholarships Foundation (IKY).
L.R. would like to thank the Department of Applied Science and Technology of the Polytechnic University of Turin, and in particular Laura Andrianopoli and Francesco Raffa, for financial support.

\appendix

\section{Useful formulas in quadratic MAG and cosmological aspects}\label{appa}

In this appendix we collect some formulas useful for the study of quadratic MAG and cosmological aspects of the latter that we have pursued in the main text.

\subsection{Hypermomentum and Palatini tensor contractions}\label{hypandpalacontr}

Considering the cosmological setup described in Section \ref{tb} and the hypermomentum in eq. \eqref{hypermomform}, we find the following contractions:
\begin{equation}
\begin{split}
& h^{\alpha\mu}\Delta_{\alpha\mu\nu}=(n-1)\phi u_{\nu} \,, \\
& h^{\alpha\nu}\Delta_{\alpha\mu\nu}=(n-1)\chi u_{\mu} \,, \\
& h^{\mu\nu}\Delta_{\alpha\mu\nu}=(n-1)\psi u_{\alpha} \,, \\
& \epsilon^{\alpha\mu\nu\lambda}\Delta_{\alpha\mu\nu}=-6 u^{\lambda} \zeta \delta^n_4 \,, \\
& u^{\alpha}u^{\mu}u^{\nu}\Delta_{\alpha\mu\nu}=-\omega \,, 
\end{split}
\end{equation}
that is
\begin{equation}
\begin{split}
& \phi=-\frac{1}{(n-1)}h^{\alpha\mu}u^{\nu}\Delta_{\alpha\mu\nu} \,, \\
& \chi=-\frac{1}{(n-1)}h^{\alpha\nu}u^{\mu}\Delta_{\alpha\mu\nu} \,, \\
& \psi=-\frac{1}{(n-1)}h^{\mu\nu}u^{\alpha}\Delta_{\alpha\mu\nu} \,, \\
& \zeta=\frac{1}{6}\epsilon^{\alpha\mu\nu\lambda}\Delta_{\alpha\mu\nu}u_{\lambda}\delta^n_4 \,, \\
& \omega = - u^{\alpha}u^{\mu}u^{\nu}\Delta_{\alpha\mu\nu} \,.
\end{split}
\end{equation}
On the other hand, regarding the Palatini tensor, whose definition is given in \eqref{palatinidefin}, we get
\begin{equation}\label{palatinicosm}
\begin{split}
P_{\alpha\mu\nu} & =\left[\frac{(n-3)}{2}A+2(n-2)\Phi -\frac{C}{2}\right] u_{\alpha}h_{\mu\nu}+\left[\frac{(n-2)}{2}B-\frac{(n-1)}{2}A -2(n-2)\Phi -\frac{C}{2}\right] u_{\mu}h_{\alpha\nu} \\
& -\frac{B}{2}u_{\nu}h_{\mu\alpha}-\frac{(n-1)}{2}Bu_{\alpha}u_{\mu}u_{\nu}-2\epsilon_{\alpha\mu\nu\rho}u^{\rho}P \delta^n_4 \,.
\end{split}
\end{equation}
Then, we have
\begin{equation}
\begin{split}
& h^{\alpha\mu}P_{\alpha\mu\nu}=-\frac{(n-1)}{2}B u_{\nu} \,, \\
& h^{\alpha\nu}P_{\alpha\mu\nu}=(n-1)\left[\frac{(n-2)}{2}B-\frac{(n-1)}{2}A -2(n-2)\Phi -\frac{C}{2}\right] u_{\mu} \,, \\
& h^{\mu\nu}P_{\alpha\mu\nu}=(n-1)\left[\frac{(n-3)}{2}A+2(n-2)\Phi -\frac{C}{2}\right] u_{\alpha} \,, \\
& \epsilon^{\alpha\mu\nu\lambda}P_{\alpha\mu\nu}= 12 P u^{\lambda} \delta^n_4 \,, \\
& u^{\alpha}u^{\mu}u^{\nu}P_{\alpha\mu\nu}= \frac{(n-1)}{2}B
\end{split}
\end{equation}
as fundamental contractions.

\subsection{Variation of the quadratic non-metricity terms with respect to the metric}\label{usefulvar}

Let us report here some calculations regarding the variation of the quadratic pure non-metricity terms in the general quadratic theory \eqref{genact}, as it is slightly more complicated with respect to the variation of the other quadratic terms.
First of all, note that the pure non-metricity quadratic combinations in the general quadratic model can be written as
\begin{equation}
\begin{split}
\mathcal{L}_{Q} & =	a_{1}Q_{\alpha\mu\nu}Q^{\alpha\mu\nu} + a_{2}Q_{\alpha\mu\nu}Q^{\mu\nu\alpha} +
a_{3}Q_{\mu}Q^{\mu}+ a_{4}q_{\mu}q^{\mu}+ a_{5}Q_{\mu}q^{\mu} \\
& = {Q_{\alpha}}^{\beta\gamma}{Q_{\sigma}}^{\lambda\rho}\Big( a_{1}g^{\sigma\alpha}g_{\lambda\beta}g_{\gamma\rho}+a_{2}\delta_{\beta}^{\sigma}\delta_{\rho}^{\alpha}g_{\gamma\lambda}+a_{3}g_{\beta\gamma}g_{\lambda\rho}g^{\sigma\alpha}+a_{4}\delta_{\lambda}^{\sigma}\delta_{\beta}^{\alpha}g_{\gamma\rho}+a_{5}\delta_{\lambda}^{\sigma}\delta_{\rho}^{\alpha}g_{\beta\gamma}\Big) \,.
\end{split}
\end{equation}
Then, defining
\beq
{D^{\sigma\alpha}}_{\lambda\beta\gamma\rho}:=\Big( a_{1}g^{\sigma\alpha}g_{\lambda\beta}g_{\gamma\rho}+a_{2}\delta_{\beta}^{\sigma}\delta_{\rho}^{\alpha}g_{\gamma\lambda}+a_{3}g_{\beta\gamma}g_{\lambda\rho}g^{\sigma\alpha}+a_{4}\delta_{\lambda}^{\sigma}\delta_{\beta}^{\alpha}g_{\gamma\rho}+a_{5}\delta_{\lambda}^{\sigma}\delta_{\rho}^{\alpha}g_{\beta\gamma}\Big) \,,
\eeq
we have
\beq
\mathcal{L}_{Q}={Q_{\alpha}}^{\beta\gamma}{Q_{\sigma}}^{\lambda\rho}{D^{\sigma\alpha}}_{\lambda\beta\gamma\rho} \,,
\eeq
and, introducing the quantities
\begin{equation}
\begin{split}
& ^{(1)}{\Omega^{\alpha}}_{\beta\gamma}:={Q_{\sigma}}^{\lambda\rho}{D^{\sigma\alpha}}_{\lambda\beta\gamma\rho}=a_{1}{Q^{\alpha}}_{\beta\gamma}+a_{2}{Q_{\beta\gamma}}^{\alpha}+a_{3}Q^{\alpha}g_{\beta\gamma}+a_{4}\delta^{\alpha}_{\beta}q_{\gamma}+a_{5}q^{\alpha}g_{\beta\gamma} \,, \\
& ^{(2)}{\Omega^{\sigma}}_{\lambda\rho}:={Q_{\alpha}}^{\beta\gamma}{D^{\sigma\alpha}}_{\lambda\beta\gamma\rho}=a_{1}{Q^{\sigma}}_{\lambda\rho}+a_{2}{Q_{\rho\lambda}}^{\sigma}+a_{3}Q^{\sigma}g_{\lambda\rho}+a_{4}\delta^{\sigma}_{\lambda}q_{\rho}+a_{5}\delta^{\sigma}_{\lambda}Q_{\rho} \,, 
\end{split}
\end{equation}
it follows that
\beq
\mathcal{L}_{Q} = ^{(1)}{\Omega^{\alpha}}_{\beta\gamma} {Q_{\alpha}}^{\beta\gamma}= ^{(2)}{\Omega^{\sigma}}_{\lambda\rho} {Q_{\sigma}}^{\lambda\rho} \,.
\eeq
Accordingly, we may express the variation with respect to the metric as
\beq
\delta_{g}(\sqrt{-g}\mathcal{L}_{Q})=-\frac{1}{2}\sqrt{-g}\mathcal{L}_{Q}g_{\mu\nu}\delta g^{\mu\nu}+\sqrt{-g} \delta_g \mathcal{L}_{Q} \,,
\eeq
and, after some algebra, we find
\begin{equation}
\begin{split}
\delta_{g}(\sqrt{-g}\mathcal{L}_{Q})& =\delta g^{\mu\nu}\sqrt{-g}\Big[ -\frac{1}{2}\mathcal{L}_{Q}g_{\mu\nu}+\frac{1}{\sqrt{-g}}\hat{\nabla}_{\alpha}(\sqrt{-g}{W^{\alpha}}_{(\mu\nu)})+a_{1}(Q_{\mu\alpha\beta}{Q_{\nu}}^{\alpha\beta}-2 Q_{\alpha\beta\mu}{Q^{\alpha\beta}}_{\nu})-a_{2}Q_{\alpha\beta(\mu}{Q^{\beta\alpha}}_{\nu)} \\
& + a_{3}(Q_{\mu}Q_{\nu}-2 Q^{\alpha}Q_{\alpha\mu\nu})-a_{4}q_{\mu}q_{\nu}-a_{5}q^{\alpha}Q_{\alpha\mu\nu} \Big] \,,
\end{split}
\end{equation}
where $\hat{\nabla}$ is defined in \eqref{hatnabla} and ${W^{\alpha}}_{(\mu\nu)}$ is given by \eqref{Wtensor}.

\subsection{Advantageous relations for FLRW cosmology of quadratic MAG}\label{usefulrelFLRW}

Let us now collect some important relations we used throughout our study of the FLRW cosmology of quadratic MAG. 
First, let us compute the form of the quadratic Lagrangian
\begin{equation}\label{L2}
\begin{split}
\mathcal{L}_{2} & := b_{1}S_{\alpha\mu\nu}S^{\alpha\mu\nu} + b_{2}S_{\alpha\mu\nu}S^{\mu\nu\alpha} +
b_{3}S_{\mu}S^{\mu} \nonumber \\
& + a_{1}Q_{\alpha\mu\nu}Q^{\alpha\mu\nu} + a_{2}Q_{\alpha\mu\nu}Q^{\mu\nu\alpha} + a_{3}Q_{\mu}Q^{\mu}+ a_{4}q_{\mu}q^{\mu}+ a_{5}Q_{\mu}q^{\mu} \\
& + c_{1}Q_{\alpha\mu\nu}S^{\alpha\mu\nu}+ c_{2}Q_{\mu}S^{\mu} + c_{3}q_{\mu}S^{\mu} \,.
\end{split}
\end{equation}
Using the cosmological forms of torsion and non-metricity tensors given by \eqref{tornonmetFLRW}, after some straightforward algebra we find
\begin{equation}
\begin{split}
\mathcal{L}_{2} & = -(n-1)\Big[ a_{1}+(n-1)a_{3} \Big] A^{2}-\frac{(n-1)}{4}\Big[ 2 a_{1}+a_{2}+(n-1)a_{4}\Big] B^{2}-(a_{1}+a_{2}+a_{3}+a_{4}+a_{5})C^{2} \\
& -(n-1)\Big[ a_{2}+\frac{(n-1)}{2}a_{5} \Big]A B+(n-1)(2 a_{3}+ a_{5}) AC +(n-1)\Big( a_{4}+\frac{1}{2}a_{5}\Big)B C \\
& -(n-1)\Big[ 2 b_{1}-b_{2}+(n-1) b_{3} \Big] \Phi^{2}+ 6(b_{1}+b_{2})P^{2} \delta^n_4 \\
& -(n-1)\Big[ c_{1}+(n-1) c_{2} \Big] A \Phi+\frac{(n-1)}{2} \Big[ c_{1}-(n-1) c_{3} \Big] B \Phi +(n-1)(c_{2}+c_{3})C \Phi \,.
\end{split}
\end{equation}
Notice that all $A,B,C$ and $\Phi$ couple among each other, while the pseudo-scalar mode $P$ appears only as $P^{2}$. Of course there is no surprise here, since $P$ is a pseudo-scalar, the total Lagrangian $\mathcal{L}_{2}$ must be a scalar, and the only combination in $P$ that can yield a scalar result is $P^{2}$ (i.e., indeed, pseudo-scalar $\times$ pseudo-scalar).
Here we also compute the following quantities:
\begin{equation}
\begin{split}
& {W^{\alpha}}_{\mu\nu}u^{\mu}u^{\nu}=2(a_{1}+a_{2}+a_{3}+a_{4}+a_{5})Cu^{\alpha}-(n-1)(2 a_{3}+ a_{5})A u^{\alpha} -(n-1)(a_{5}+2 a_{4})\frac{B}{2}u^{\alpha} \,, \\
& {\Pi^{\alpha}}_{\mu\nu}u^{\mu}u^{\nu}=-(n-1)(c_{2}+c_{3})\Phi u^{\alpha} \,, \\
& {W^{\alpha}}_{(\mu\alpha)}=\Big[ a_{2}+2 a_{3}+\frac{(n+1)}{2}a_{5}\Big] Q_{\mu}+\Big[ 2 a_{1}+ a_{2}+ 2 a_{3}+a_{5}+(n+1)a_{4} \Big] q_{\mu} \\
& \phantom{{W^{\alpha}}_{(\mu\alpha)}} = \frac{1}{2} (n-1) \left[ 2 a_2 + 4 a_3 + (n+1) a_5 \right] A u_\mu + \frac{1}{2} (n-1) \left[ 2 a_1 + a_2 + a_5 + (n+1) a_4 \right] B u_\mu \\
& \phantom{{W^{\alpha}}_{(\mu\alpha)}} + \frac{1}{2} \left[ -4 a_1 - 4 a_2 -4a_3 - 2 (n+1) a_4 - a_5 (n+3) \right] C u_\mu \,, \\  
& {\Pi^{\alpha}}_{(\mu\alpha)}=\Big[ -\frac{c_{1}}{2}+c_{2}+\frac{(n+1)}{2}c_{3} \Big] S_{\mu} = \left[ - c_1 + 2 c_2 + (n+1) c_3 \right] \Phi u_\mu \,, \\
& W_{\alpha\mu\nu}u^{\alpha}u^{\mu}u^{\nu}=2(a_{1}+a_{2})Q_{\alpha\mu\nu}u^{\alpha}u^{\mu}u^{\nu}-(2 a_{3}+a_{5})Q_{\mu}u^{\mu}-(a_{5}+ 2 a_{4})q_{\mu}u^{\mu} \\
& \phantom{W_{\alpha\mu\nu}u^{\alpha}u^{\mu}u^{\nu}} = -2(a_{1}+a_{2}+a_{3}+a_{4}+a_{5})C+(n-1)(2 a_{3}+a_{5})A+(n-1)(a_{5}+2 a_{4})\frac{B}{2} \,, \\
& \Pi_{\alpha\mu\nu}u^{\alpha}u^{\mu}u^{\nu}=-(c_{2}+c_{3})S_{\mu}u^{\mu}=(n-1)(c_{2}+c_{3}) \Phi \,.
\end{split}
\end{equation}
Besides, given \eqref{Ruu} and specializing to the FLRW cosmology case we considered in the main text (i.e., using \eqref{tornonmetFLRW}), we find
\begin{equation}
\begin{split}
(A_{\mu\nu}+B_{\mu\nu}+C_{\mu\nu})u^{\mu}u^{\nu} & =(n-1)\Big[ a_{1}+(n-1)a_{3}\Big] A^{2}-\frac{1}{4}\Big[ (2 a_{1}+a_{2})+(n-1)a_{4} \Big] (n-1) B^{2} \\
& -(a_{1} + a_{2}+a_{3}+a_{4}+a_{5})C^{2} +(n-1)\Big( a_{4}+\frac{a_{5}}{2}\Big)B C \\
& +(n-1)\Big[ 2 b_{1}-b_{2}+(n-1)b_{3}\Big] \Phi^{2}+(n-1)\Big[ c_{1}+(n-1)c_{2}\Big] \Phi A \,.
\end{split}
\end{equation}
Note that there are no $AC, AB, \Phi B, \Phi C$ terms in the equation above, that is $A$ and $\Phi$ couple only among themselves and so do $B$ and $C$. Observe also that here $P$ does not appear.
Moreover, after some long calculations we find
\begin{equation}
\begin{split}
-2({\Pi^{\alpha}}_{(\mu\nu)}+{W^{\alpha}}_{(\mu\nu)})u^{\mu}\nabla_{\alpha}u^{\nu} & = (n-1)(H-Y)\Big[ \gamma_{1} B+2\gamma_{2}A+ 2 \gamma_{3}\Phi\Big] \\
& +\frac{(n-1)}{2}C \Big[ (a_{5}+2 a_{4})B+2(2 a_{3}+a_{5})A+2(c_{2}+c_{3})\Phi \Big] \\
& -2(a_{1}+a_{2}+a_{3}+a_{4}+a_{5})C^{2}-(n-1)(2 a_{4}+a_{5})(H-Y) C \,,
\end{split}
\end{equation}
where 
\beq\label{gammacoeff}
\gamma_{1}=2 a_{1}+a_{2}+(n-1) a_{4} \,, \quad  \gamma_{2}=a_{2}+\frac{(n-1)}{2}a_{5} \,, \quad \gamma_{3}=\frac{1}{2}\Big[ -c_{1}+(n-1)c_{3} \Big] \,.
\eeq
Gathering all the above results and combining them with \eqref{Ruuexpr}, we finally arrive at
\begin{equation}
\begin{split}
R_{\mu\nu}u^{\mu}u^{\nu} & =\frac{\kappa}{(n-2)}	\Big( T+(n-2)T_{\mu\nu}u^{\mu}u^{\nu}\Big) \\
& - \Big \lbrace (n-1)\Big[ a_{1}+(n-1)a_{3}\Big] A^{2}-\frac{1}{4}\Big[ (2 a_{1}+a_{2})+(n-1)a_{4} \Big] (n-1) B^{2} \\
& -(a_{1} + a_{2}+a_{3}+a_{4}+a_{5})C^{2} +(n-1)\Big( a_{4}+\frac{a_{5}}{2}\Big)B C \\
& +(n-1)\Big[ 2 b_{1}-b_{2}+(n-1)b_{3}\Big] \Phi^{2}+(n-1)\Big[ c_{1}+(n-1)c_{2}\Big] \Phi A \Big \rbrace \\
& +(n-1)(H-Y)\Big[ \gamma_{1} B+2\gamma_{2}A+ 2 \gamma_{3}\Phi\Big]+\frac{(n-1)}{2}C \Big[ (a_{5}+2 a_{4})B+2(2 a_{3}+a_{5})A+2(c_{2}+c_{3})\Phi \Big] \\
& -2(a_{1}+a_{2}+a_{3}+a_{4}+a_{5})C^{2}-(n-1)(2 a_{4}+a_{5})(H-Y) C \\
& +\dot{l}+(n-1)H l \,,
\end{split}
\end{equation}
where 
\beq\label{elle}
l=\beta_{1}A+\beta_{2}B+\beta_{3}C+\beta_{4}\Phi \,,
\eeq
with
\beq\label{betacoeffs}
\begin{split}
& \beta_{1} = \left(\frac{n-1}{n-2}\right) \left[ 2 a_1 + 4 a_3 - a_5 (n-3) \right] \,, \\
& \beta_{2} = \left(\frac{n-1}{n-2}\right) \left[ a_2 - a_4 (n-3) + a_5 \right] \,, \\
& \beta_{3} = \frac{1}{\left(n-2\right)} \left[ 2 a_1 (n-3)+2 a_2 (n-3) - 4 a_3 + 2 a_4 (n-3) + a_5 (n-5) \right] \,, \\
& \beta_{4} = \left(\frac{n-1}{n-2}\right) \left[ c_1 + 2 c_2 - c_3 (n-3) \right] \,.
\end{split}
\eeq
All these formulas are particularly useful to reproduce the results we have presented in Section \ref{cosmquadrMAG}.  

\subsection{Coefficients in quadratic MAG cosmology}\label{coeffsMAG}

For the sake of completeness, in the following we report the explicit expressions of the coefficients $\alpha_{ij}$ appearing in eqs. \eqref{e1}-\eqref{e4} in terms of the $a$'s, $b$'s and $c$'s. They read
\begin{equation}
\begin{split}
& \alpha_{11} = 2(n-1)(2a_{3}+a_{5}) \,, \quad \alpha_{12} = (n-1)\Big( \frac{1}{2}+a_{5}+2 a_{4} \Big) \,, \\
& \alpha_{13} = -4(a_{1}+a_{3}+a_{4}+a_{5}) \,, \quad \alpha_{14} = 2(n-1)(c_{2}+c_{3}) \,, \\
& \alpha_{21} = \frac{(n-3)}{2}+(n-1)a_{5} \,, \quad \alpha_{22} = 2 a_{1}+(n-1)a_{4} \,, \\
& \alpha_{23} = -\Big(\frac{1}{2}+2a_{4}+a_{5}\Big) \,, \quad \alpha_{24} = 2(n-2)+(n-1)c_{3} \,, \\
& \alpha_{31} = 4 a_{1}+4(n-1)a_{3}-\frac{(n-1)}{2}c_{2} \,, \quad \alpha_{32} = (n-1) a_{5}-\frac{(n-1)}{4}c_{3}  -\frac{1}{2} \,, \\
& \alpha_{33} = \frac{1}{2}(c_{2}+c_{3})-4 a_{3}-2 a_{5} \,, \quad \alpha_{34} = 2(n-1)c_{2}-2 b_{1}-(n-1) b_{3} \,, \\
& \alpha_{41} = (n-1) \left( -\frac{1}{2}+a_{5}+\frac{c_{2}}{2} \right) \,, \quad \alpha_{42} = \frac{(n-2)}{2}+2 a_{1}+(n-1) a_{4}+\frac{(n-1)}{4}c_{3} \,, \\
& \alpha_{43} = -\left[ \frac{1}{2}+ 2 a_{4}+a_{5}+\frac{c_{2}+c_{3}}{2} \right] \,, \alpha_{44} = -2(n-2)+ 2 b_{1}+(n-1)c_{3}+(n-1)b_{3} \,.
\end{split}
\end{equation}
These are the elements of the matrix $M$ we have introduced in Section \ref{cosmquadrMAG}.


\end{document}